\documentclass[aps,prl,showpacs,twocolumn,floats,epsfig,pdflatex]{revtex4}
\usepackage{amssymb}
\usepackage{amsbsy}
\usepackage{amsmath}
\usepackage{epsfig}
\usepackage{graphicx}
\newcommand\beq{\begin{equation}}
\newcommand\eeq{\end{equation}}
\newcommand\bea{\begin{eqnarray}}
\newcommand\eea{\end{eqnarray}}

\begin{document}

\title {Statistics of work distribution in periodically driven closed quantum systems}

\author{Anirban Dutta, Arnab Das, and K. Sengupta}

\affiliation{Theoretical Physics Department, Indian Association for
the Cultivation of Science, Jadavpur, Kolkata 700 032, India.}
\date{\today}

\begin{abstract}
We study the statistics of the work distribution $P(w)$ in a
$d-$dimensional closed quantum system with linear dimension $L$
subjected to a periodic drive with frequency $\omega_0$. We show
that the corresponding rate function $I(w)= -\ln[P(w)]/L^d$ after a
drive period satisfies an universal lower bound $I(0)\ge n_d$ and
has a zero at $w/N=Q$, where $n_d$ and $Q$ are the defect density
and the residual energy generated during the drive and $N$ denotes
the total number of sites. We supplement our results by calculating
$I(w)$ for a class of $d$-dimensional integrable models and show
that $I(w)$ has oscillatory dependence on $\omega_0$ originating
from Stuckelberg interference generated due to double passage
through critical point/region during the drive. We suggest
experiments to test our theory.
\end{abstract}


\maketitle

The study of non-equilibrium dynamics of closed quantum systems has
gained tremendous momentum in recent years due to available
experimental test beds in the form of ultracold atom systems
\cite{rev1,rev2,rev3}. Such cold atoms serve as near perfect
emulators of model Hamiltonians such as the Ising or the
Bose-Hubbard models \cite{isingref,boseref}; in addition, they offer
real time tunability of the parameters of the emulated Hamiltonians
\cite{bloch1,bakr1,venga1}. Consequently, they form perfect
experimental platforms for studying non-equilibrium dynamics of
these Hamiltonians near their quantum critical points. The initial
focus of theoretical studies in this direction has been on sudden
quench \cite{sengupta1,calabrese1} and ramp protocols
\cite{anatoli1,sengupta2, sen1, dutta1}. However, later studies have
also focussed on periodic protocols which involve multiple passage
through these critical points \cite{sengupta3,das1,das2} leading to
dynamic freezing \cite{das1, sengupta3} and novel steady states
\cite{das2}.

One of the quantities of interest in a many-body system driven out
of equilibrium is its statistics of work distribution
\cite{hugo1,hugo2,silva1,silva2,heyl1,esposito1}. For thermodynamic
systems in equilibrium, such a distribution follows the large
deviation principle (LDP), namely $P(w) \sim \exp[-L^d I(w)]$, where
$I(w)$ is the rate function characterizing the decay rate of $P(w)$
from its peak value which occurs at $w=\langle w\rangle$, where
$\langle w \rangle $ is the average work done, and $L$ is the linear
dimension of the system. LDP is also shown to be valid for a large
class of classical (quantum) non-equilibrium systems where the
dynamics can be described by Markov processes (quench or ramp
protocols)\cite{hugo2, silva1,silva2}. The latter works on quantum
systems made general arguments about features of $I(w)$ for
quench/ramp protocols and computed it explicitly for a class of
one-dimensional (1D) integrable models \cite{silva1,silva2}. Such
studies, however,  were never extended for periodic protocols beyond
two-level systems \cite{esposito1}; in particular, the effect of
multiple passage through critical points due to the drive on $P(w)$
has not been studied.

In this work, we study the statistics of the work distribution
$P(w)$ in a $d-$dimensional closed quantum system with linear
dimension $L$ subjected to a periodic drive with frequency
$\omega_0$. We provide formally exact expression of the moment
generating function, $G(u) =\int dw P(w) \exp[-u w]$, for such a
system after a drive period. Using the expression of $G(u)$ and
basic elements of large deviation theory, we show that the
corresponding rate function $I(w)= -\ln[P(w)]/L^d$ satisfies an
universal lower bound $I(0)\ge n_d$, where $n_d$ is the excitation
(defect) density generated during the drive. We also show that for
any periodic protocol $I(Q)=0$ where $Q$ is the residual energy. We
supplement our results by explicit calculation of $I(w)$ for a class
of integrable models in $d$-dimensions. Specific examples of these
models include the Ising and XY models in $d=1$ and the Kitaev model
in $d=2$. We show that $I(w)$ has an non-monotonic dependence on the
drive frequency $\omega_0$ which originates from the Stuckelberg
interference generated during multiple passage of these systems
through quantum critical points or lines during the drive. We
suggest concrete experiments to test our theory. To the best of our
knowledge, the existence of a universal lower bound for $I(0)$
linking it to an experimentally measurable quantity $n_d$ has never
been shown for a generic periodically driven quantum system; also,
the relevance of quantum interference for statistics of work
distribution of a closed quantum system has not been pointed out.
Our work aims to fill up these major gaps in the existing
literature.

Consider a time-dependent Hamiltonian $H[\lambda(t)] \equiv H$ where
the parameter $\lambda(t)$ has periodic time dependence with a
characteristics frequency $\omega_0$. At $t=0$,
$\lambda(0)=\lambda_0$ and $\lambda(t)$ returns to its starting
value $\lambda_0$ after a drive cycle at $t_f = T_0=2\pi /\omega_0$.
We denote the eigenstates and eigenenergies of $H$ at $t=0,t_f$ by
$|\alpha\rangle$ and $E_{\alpha}$; they obey $H[\lambda_0] |\alpha
\rangle= E_{\alpha} |\alpha\rangle$. For such a quantum system in
the ground state ($|0\rangle$) of $H$ at $t=0$, $w$ can take a set
of values $E_{\alpha}-E_0$; thus $P(w)$ is obtained by summing over
probabilities of $w$ being equal to $E_{\alpha}-E_0$ for all
$\alpha$:
\begin{eqnarray}
P(w) &=& \sum_{\alpha}P_{\alpha}(w) = \sum_{\alpha} P_{|0\rangle \to
|\alpha\rangle} \delta (w-E_{\alpha} +E_0), \label{probdist}
\end{eqnarray}
where $P_{|0\rangle \to |\alpha\rangle}$ denotes the probability
that the system reaches the state $|\alpha\rangle$ at the end of a
drive cycle; it can be expressed in terms of the evolution operator
$S$ as
\begin{eqnarray}
P_{|0\rangle \to |\alpha\rangle} &=& |\langle \alpha |S|0\rangle|^2,
\quad S= T_t e^{-\frac{i}{\hbar} \int_0^{t_f} H[\lambda(t')]dt'},
\label{eveq}
\end{eqnarray}
where $T_t$ denotes time ordering. We note that one can write
$S|0\rangle = |\psi(t_f)\rangle = \sum_{\alpha} c_{\alpha}
|\alpha\rangle$, where $c_{\alpha}$ denotes the wavefunction overlap
between $|\psi(t_f)\rangle$ with the eigenstate $|\alpha\rangle$;
they satisfy $\sum_{\alpha} |c_{\alpha}|^2 =1$. The probability of
finding the system in the ground (starting) state after a drive
cycle is $|c_0|^2$. The total energy absorbed during such a drive
(residual energy) is thus given by $Q = L^{-d} \sum_{\alpha}
(E_{\alpha}-E_0) |c_{\alpha}|^2$.

The moment generating function of $P(w)$, given by its Laplace
transform, can be written as \cite{silva1}
\begin{eqnarray}
G(u)= \int dw \exp[-w u] P(w) = \sum_{\alpha} P_{|0\rangle \to
|\alpha\rangle} e^{-(E_{\alpha}-E_0)u}. \nonumber\ \label{geneq0}
\end{eqnarray}
Using Eq.\ \ref{eveq} and the relations $H(t_f)=H(0)\equiv
H(\lambda_0)$ and $S|0\rangle = \sum_{\alpha} c_{\alpha}
|\alpha\rangle$, one can express $G(u)$ as
\begin{eqnarray}
G(u) &=& |c_0|^2 +\sum_{\alpha \ne 0} |c_{\alpha}|^2
e^{-(E_{\alpha}-E_0)u}. \label{geneq2}
\end{eqnarray}

Next we show that $G(u)$ obtained in Eq.\ \ref{geneq2} satisfies
LDP, {\it i.e.}, $G(u)=\exp[-L^d f(u)]$, where $f(u)$ is a concave
function \cite{hugo1}. To do this, we express $f(u)$ as (Eq.\
\ref{geneq2})
\begin{eqnarray}
f(u)= -L^{-d} \ln \Big[|c_0|^2 +\sum_{\alpha \ne 0} |c_{\alpha}|^2
e^{-(E_{\alpha}-E_0)u} \Big]. \label{feq1}
\end{eqnarray}
To show that $f(u)$ is a concave function, we observe that $f(0)=0$
and $f(\infty) = L^{-d} \ln[1/|c_0|^2] \ge 0$ \cite{comment0}.
Moreover,
\begin{eqnarray}
\partial_u f(u) &=& L^{-d} \frac{ \sum_{\alpha \ne 0} (E_{\alpha}-E_0)
|c_{\alpha}|^2 e^{-(E_{\alpha} -E_0) u}}{|c_0|^2 + \sum_{\alpha \ne
0} |c_{\alpha}|^2 e^{-(E_{\alpha} -E_0) u}}\label{fdereq}
\end{eqnarray}
vanishes only at $u = \infty$ since $E_{\alpha}-E_0$ are positive
definite for all $\alpha$. Further $\partial_u^2 f(u) <0$; so for
any generic closed periodically driven quantum system, $f(u)$,
computed after a drive cycle, is a concave function in $u \in
(0,\infty)$. Thus $G(u)$ (and hence $P(w)$) obeys LDP.

Since $G(u)$ obeys LDP, one can relate rate function $I(w)= -L^{-d}
\ln[P(w)]$ to $f(u)$ using the Gartner-Ellis theorem
\cite{hugo1,hugo2,comment1}:
\begin{eqnarray}
I({\bar w}) &=&  f(u[{\bar w}]) - {\bar w} u[{\bar w}], \quad
\partial_u f(u)|_{u=u[{\bar w}]} = {\bar w}, \label{gerate}
\end{eqnarray}
where ${\bar w}= w/N$ and $N$ is the number of sites. First, we use
Eq.\ \ref{gerate} to relate the zero of $I(w)$ to $Q$. To this end,
we use Eq.\ \ref{fdereq} to obtain $Q =\partial_u f(u)|_{u=0}$.
Using this observation, we note that Eq.\ \ref{gerate} admits a
solution $u[{\bar w}]=0$ for ${\bar w}=Q $ leading to $I(Q)=f(0)=0$.

Next, we obtain the universal lower bound for $I(0)$. To this end,
we observe from Eq.\ \ref{gerate} that $I(0)= f(u[0])$, where
$\partial_u f(u)|_{u=u[0]}=0$. From Eq.\ \ref{fdereq}, we find that
$u[0]=\infty$ which leads to $I(0)=f(\infty)=-L^{-d} \ln(|c_0|^2)$
\cite{comment2}. To relate $|c_0|^2$ to $n_d$, we first consider a
class of integrable systems for which $|c_0|^2 =
\prod_{j=1,N}(1-p_j)$\cite{intref,comment00}. Here $p_j$ denotes the
probability of deviation of the system from the ground state
configuration corresponding to the index $j$. The physical
interpretation of index $j$  depends on the system; for example, it
may represent either a spatial site or a momentum mode. Using this,
one sees that $I(0) = - L^{-d} \sum_j\ln(1-p_j) \ge Np/L^d$, where
$Np=\sum_j p_j$. Next, we note that a finite $p$ indicates non-zero
weight of the system in the excited state; in an integrable system
this amounts to formation of quasiparticle excitations whose number
is $Np$ and density $n_d= Np/L^d$. Thus for integrable systems $I(0)
\ge n_d$. This result can be generalized for systems at finite
temperature \cite{comment00}.

To relate $I_0$ to $n_d$ for non-integrable systems, we divide $T_0$
into ${\mathcal M}$ intervals $\Delta t= T_0/{\mathcal M}$ with
small enough $\Delta t$ so that for any interval $j$,
$H[\lambda(t)]\simeq (H[\lambda(t_j)] + H[\lambda(t_{j+1})])/2$.
This leads to a time independent Hamiltonian, $H \simeq
H[\lambda[t_j]] + \Delta t
\partial_t H(t)|_{t=t_j}/2 = H_j + \Delta t V_j $, which
describes the evolution of the system wavefunction at the $j^{\rm
th}$ step. In what follows, we denote $|n_j\rangle$ and $E_{n_j}$ to
be instantaneous eigenstates and eigenenrgies of $H_j$. For small
enough $\Delta t$, $\Delta t V_j \ll H_j$; thus one can estimate the
wavefunction evolution within any step using time-independent
perturbation theory. Such a division of $T_0$ also allows one to
write $S = \prod_{j=1,{\mathcal M}} \exp[-i (H_j+\Delta t V_j)\Delta
t/\hbar]$. The system wavefunction, after $j-1$ evolution steps, can
be written as $|\psi_j\rangle = \sum_{n_j} a_{n_j}^j |n_j\rangle$,
where $a_{n_j}^j = \langle n_j|\psi_j\rangle$. Since
$|\psi_{j+1}\rangle = \sum_{n_{j+1}} a_{n_{j+1}}^{j+1}
|n_{j+1}\rangle$ and $|n_{j+1}\rangle$ can be related to
$|n_j\rangle$ using time-independent perturbation theory,
$a_{n_j}^j$ obeys a recursion relation \cite{commentfi}
\begin{eqnarray}
a_{n_{j+1}}^{j+1} &=& (1-p_{n_j}^j/2) a_{n_j}^j + \sum_{m_j \ne n_j}
\alpha_{m_j n_j}^j a_{m_j}^j, \label{rec1}
\end{eqnarray}
where $\alpha_{m_j n_j}^j = \langle m_j |\Delta t V_j
|n_j\rangle/[E_{n_j}-E_{m_j}] + {\rm O}(\Delta t^2)$ and $p_{n_j}^j=
\sum_{m_j \ne n_j} |\alpha_{m_j n_j}^j|^2$ with $p_{0_j}^j \equiv
p_j$. Since $|0_{\mathcal M}\rangle = |0\rangle$ at the end of a
drive period, $|c_0|^2 = |\langle \psi_{\mathcal M}|0\rangle|^2=
|a_{0_{\mathcal M}}^{\mathcal M}|^2$. Using Eq.\ \ref{rec1}, one
then obtains after some algebra $|c_0|^2 = \prod_{j=0}^{{\mathcal
M}-1} (1-p'_j)$, where $p'_j = {\rm Re}(p_{1j})-|p_{1j}|^2/4$ and
$p_{1j} \simeq p_j -2 \sum_{n=1,{\mathcal M}-j} \sum_{m_j \ne 0_j}
\alpha_{m_j 0_j}^j \alpha_{m_{j-n} 0_{j-n}}^{j-n}$ \cite{commentfi}.
This leads to
\begin{eqnarray}
I(0)= -L^{-d} \sum_{j=0,{\mathcal M}-1} \ln(1-p'_j) \ge
\sum_{j=0,{\mathcal M}-1} p'_j/L^d, \label{ineq1}
\end{eqnarray}
Next, we note that $p'_j$ represents the change in the wavefunction
overlap with the instantaneous ground state during the $j^{\rm th}$
evolution step: $|a_{0_{j+1}}^{j+1}/a_{0_j}^j|^2 = (1-p'_j)$
\cite{commentfi}. Thus $p=\sum_j p'_j/{\mathcal M}$ represents the
probability of deviation of $|\psi_{{\mathcal M}}\rangle$ from
$|0\rangle$ at the end of the drive cycle. The number of resultant
excitations is $N_0={\mathcal M} p$ leading to $n_d = N_0/L^d$
\cite{commentex}. This yields $I(0)\ge n_d$ (Eq.\ \ref{ineq1}).

The relations $I(0)\ge n_d$ and $I(Q)=0$ constitute the central
results of this work. They relate $I(w)$ of a periodically driven
quantum system to physically measurable quantities $n_d$ and $Q$.
These relations are universal; they hold irrespective of the system
dimension, specific parameters of its Hamiltonian, and details of
the periodic drive protocol. These details, encoded in $H$ and
$\omega_0$, determine $n_d$ and $Q$; however they do not alter the
above-mentioned relations. We point out that for any drive protocol
(not necessarily periodic), the rate function vanishes at ${\bar w}
=\langle w \rangle/N$ \cite{silva1,silva2}; however, $\langle w
\rangle/N$ can not be related to $Q$ for such protocols. This
equality of these two quantities stems from the drive periodicity
leading to $H_f=H_i$ after a drive cycle.

\begin{figure}[t!]
\begin{center}
\includegraphics[width=0.58\columnwidth]{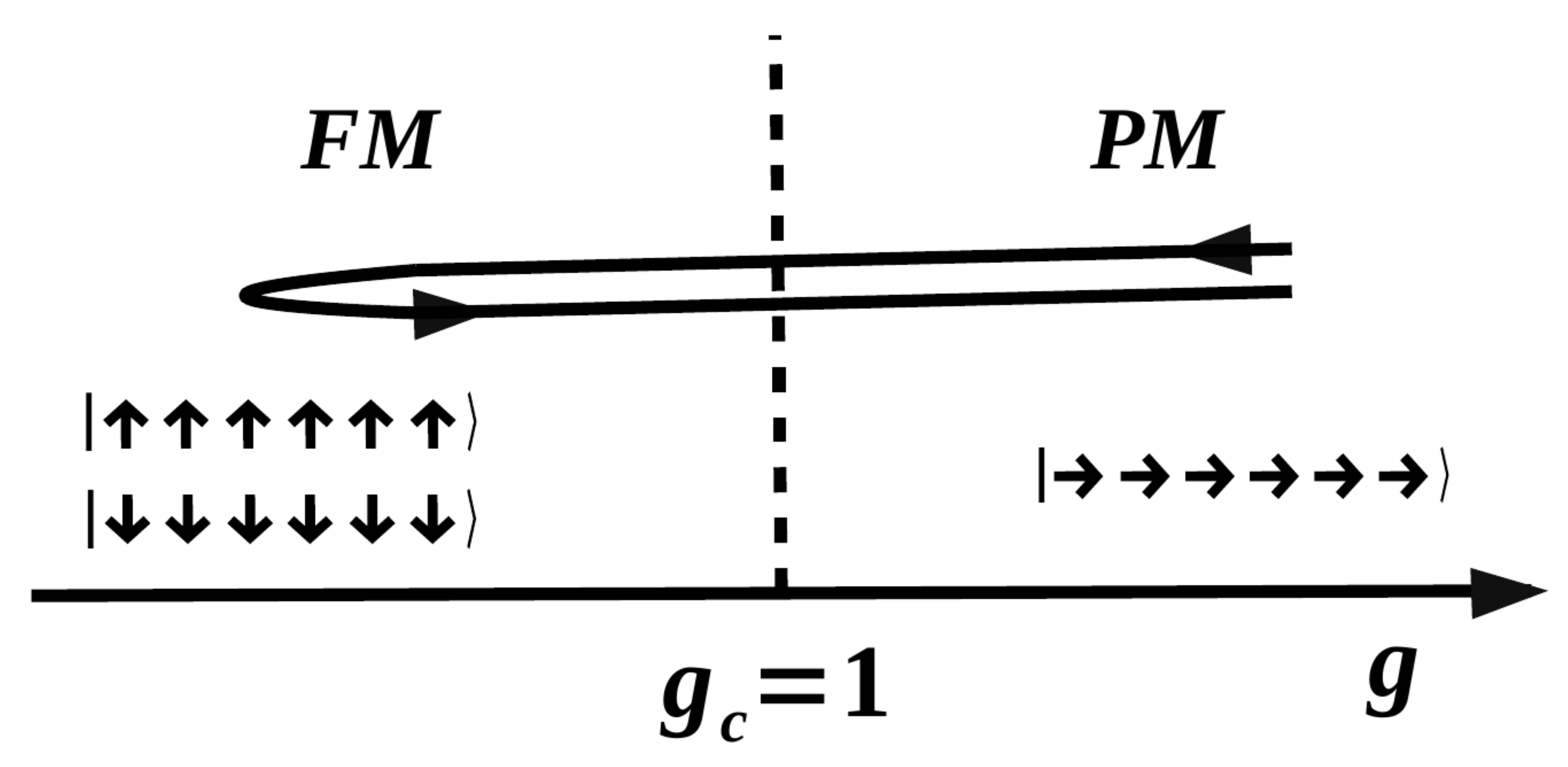}
\includegraphics[width=0.40\columnwidth]{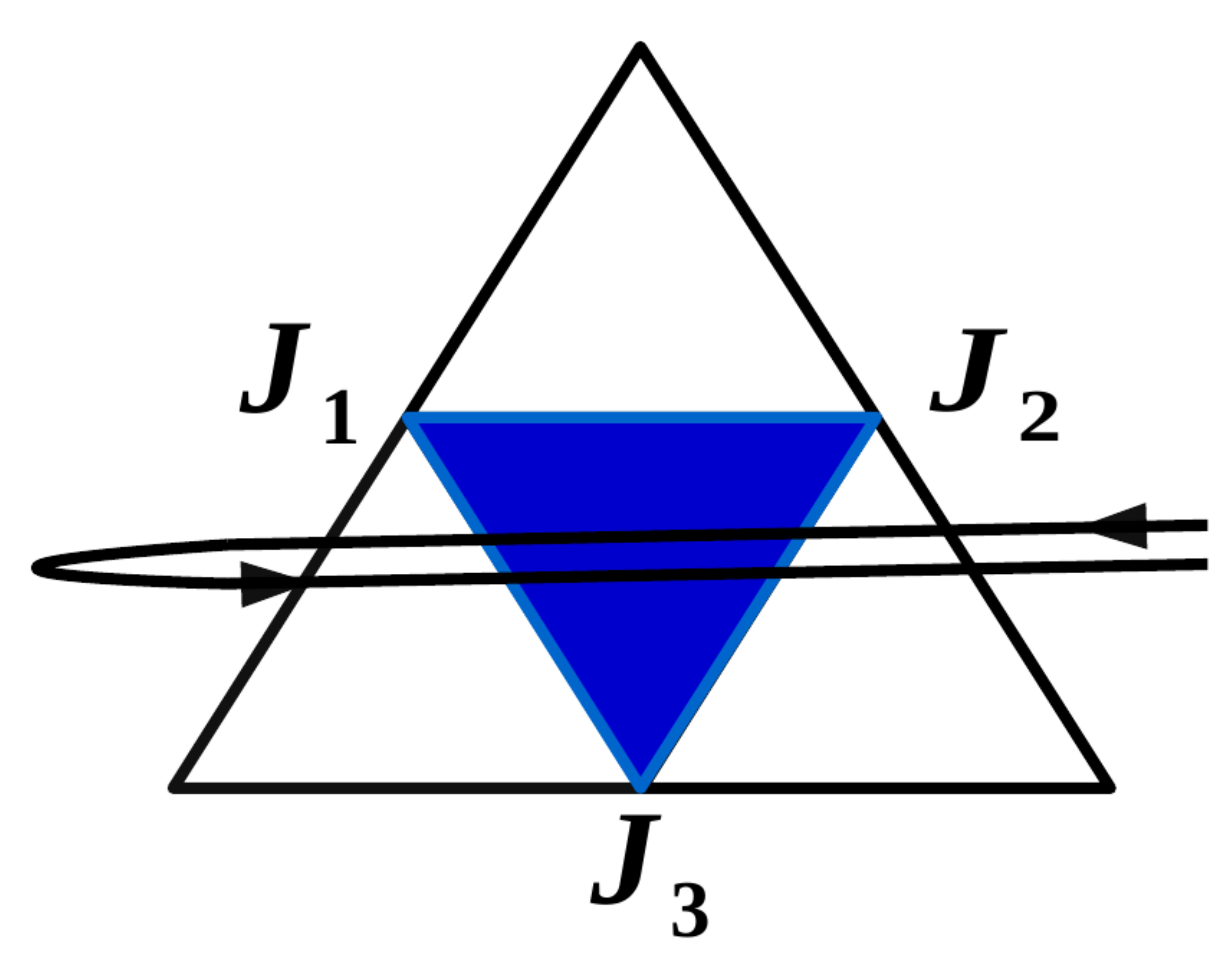}
\end{center}
\caption{(Color online) Schematic representation of the phase
diagram of the Ising (left) and the Kitaev (right) model showing
multiple passage through the critical point/region during a drive
cycle.}\label{fig1}
\end{figure}

Next, we compute $I(w)$ for $d$-dimensional integrable models
described by $ H_{\rm int}(t) = \sum_{\bf k} \psi_{\bf k}^{\dagger}
H_{\bf k}(t) \psi_k$, where $\psi_{\bf k}^{\dagger}= (c_{1 \bf
k}^{\dagger},c_{2 \bf k}^{\dagger})$ are fermionic creation
operators and
\begin{eqnarray}
H_{\bf k}(t) = \tau_3 (\lambda_1(t) -b_{\bf k}) + \tau_1 g_{\bf k}.
\label{ham1}
\end{eqnarray}
Here $\tau_3$ and $\tau_1$ denote usual Pauli matrices while $b_{\bf
k}$ and $g_{\bf k}$ are general functions of momenta. In what
follows, we consider a periodic protocol $\lambda(t)= \lambda_0
[1+\cos(\omega_0 t)]$ and compute $I(w)$ for the model at end of one
drive cycle.

In this context, we note that the Hamiltonian of Ising model is
given by $H_{\rm Ising} = -J \sum_{\langle ij\rangle} S_i^z S_j^z -h
\sum_i S_i^x$, where $J$ is the nearest-neighbor coupling between
the spins and $h$ is the transverse field. It turns out that $H_{\rm
Ising}$ reduces to $H_{\rm int}$ via Jordan-Wigner transformation
\cite{isingbook,commentkit} with $\lambda_0=g=h/J$, $b_k=\cos(k)$
and $g_k =\sin(k)$. Further, the Kitaev model, describing half
integer spins on a 2D honeycomb lattice, has the Hamiltonian
\cite{kitaevref,commentkit}
\begin{eqnarray}
H' &=& \sum_{j+\ell ={\rm even}}  J_{1} S_{j,\ell}^{x}
S_{j+1,\ell}^{x} + J_{2} S_{j,\ell}^{y} S_{j-1,\ell}^{y} +J_3
S_{j,\ell}^{z} S_{j,\ell+1}^{z}, \nonumber
\end{eqnarray}
where $J_{1,2,3}$ denote nearest neighbor coupling between the spins
and the $(j,\ell)$ describe 2D lattice coordinates. It is well-known
that $H'$ can also be mapped to $H_{\rm int}$ via a 2D Jordan Wigner
transformation \cite{kitaevref} with $\lambda_0=J_3$, $b_{\bf k}=
(J_1 \cos({\bf k}\cdot {\bf M_1}) + J_2 \cos({\bf k}\cdot {\bf
M_2}))$, and $g_{\bf k}= (J_1 \sin({\bf k}\cdot {\bf M_1}) - J_2
\sin({\bf k}\cdot {\bf M_2}))$. Here ${\bf M}_{1,2}=
(\sqrt{3}/2,+(-)3/2)$ denote the spanning vectors of the reciprocal
lattice of the model and we have set the lattice spacing to unity.

To obtain $I(w)$ for $H_{\rm int}(t)$, we first note that the
instantaneous eigenvalues of $H_{\bf k}(t)$ is given by $E_{\bf
k}[\lambda(t)] = \pm \sqrt{(\lambda(t) -b_{\bf k})^2+ g_{\bf k}^2}$.
Using this, we find that the model passes through a critical point
(line) in $d=1$ ($d=2$) where $g_{\bf k}=0$ as shown schematically
in Fig.\ \ref{fig1}. For $d=1$, the critical point is reached twice
for each cycle at $t_1= \omega_0^{-1} \arccos(b_{{k}_0}/\lambda_0-1)
$ and $t_2=2\pi/\omega_0 -t_1$, where $g_{k_0}=0$. For $d=2$, the
critical region is traversed during time windows $t_i= \omega_0^{-1}
\arccos(b_{{\bf k}_i}/\lambda_0-1)$ and $t'_i= 2\pi/\omega_0-t_i$,
where ${\bf k}_i$ satisfies $g_{{\bf k}_i}=0$.
\begin{figure}[t!]
\begin{center}
\includegraphics[width=0.55\columnwidth]{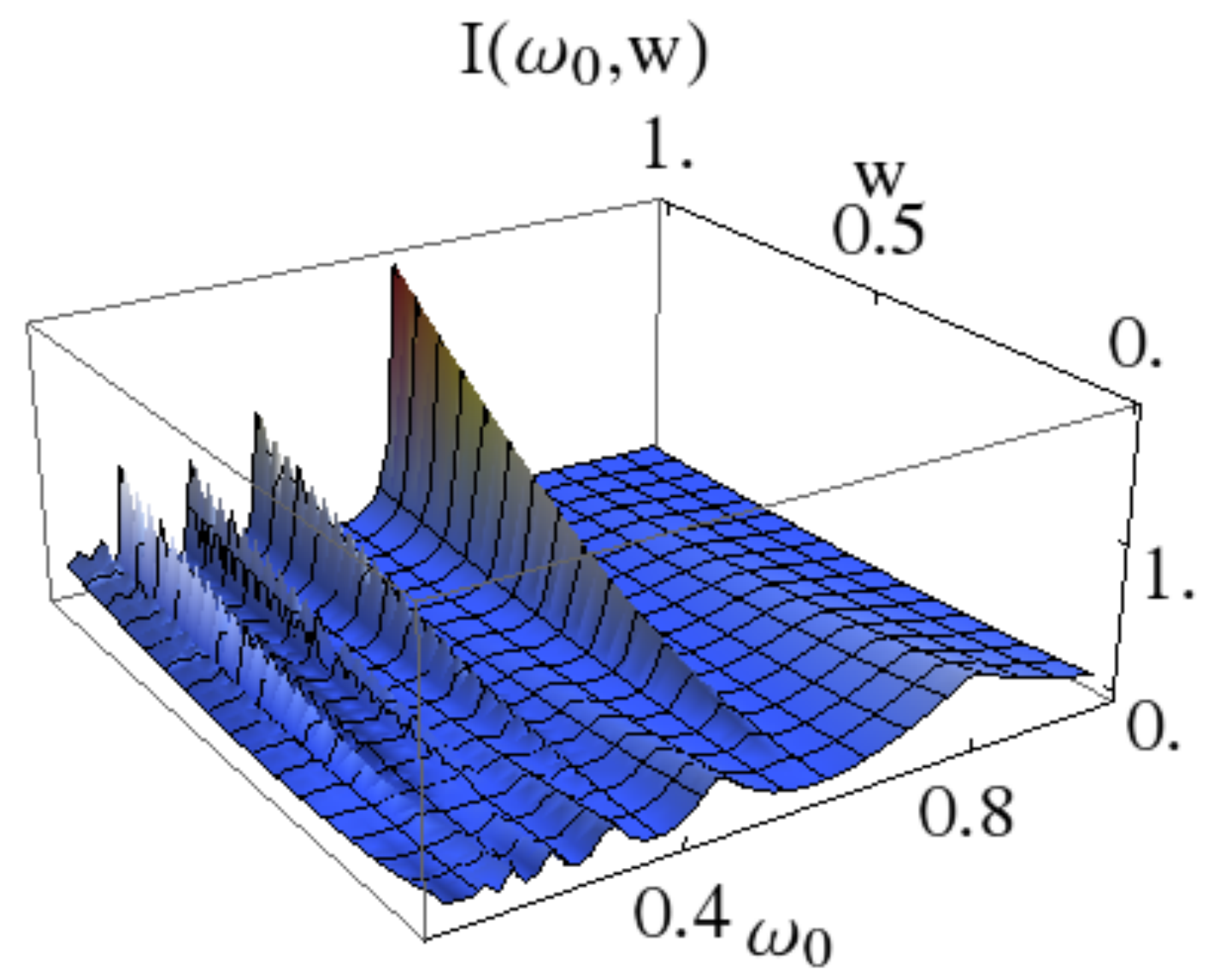}
\includegraphics[width=0.43\columnwidth]{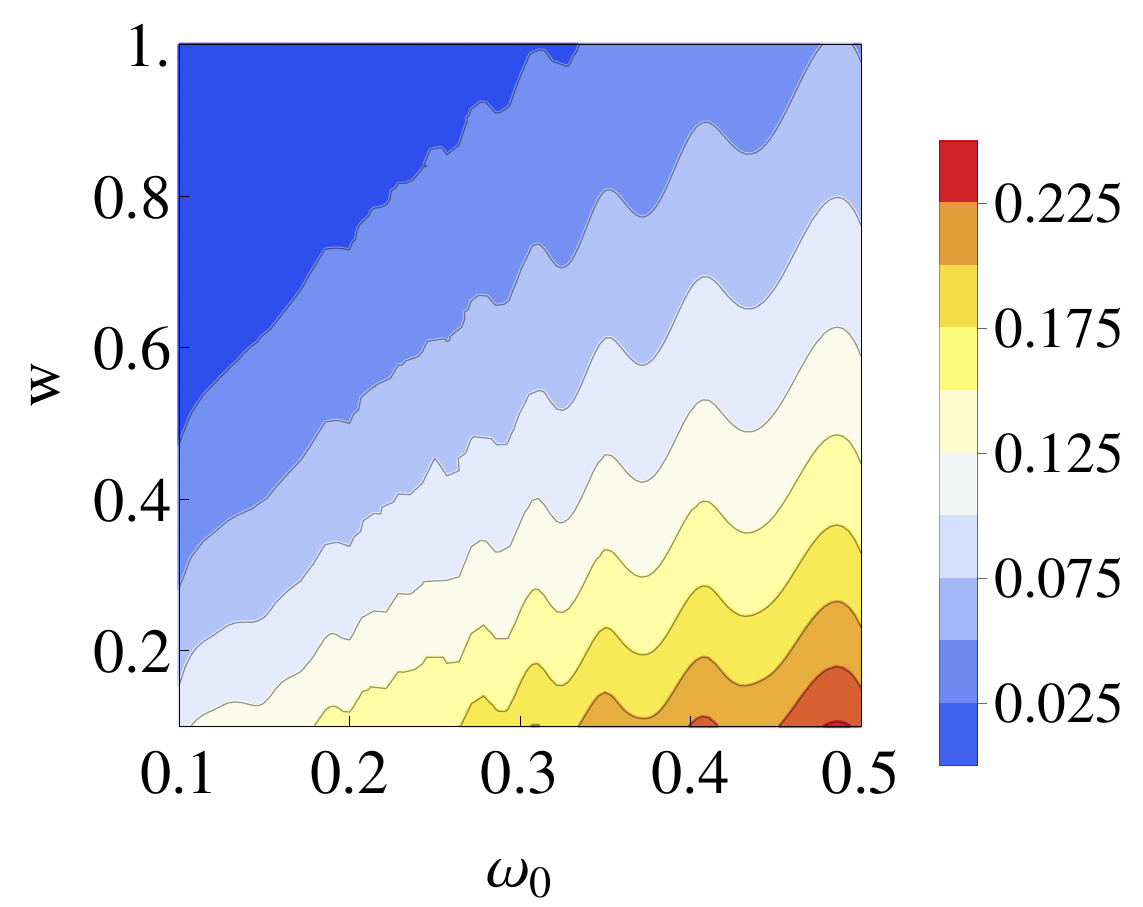}
\end{center}
\caption{(Color online) Left: Plot of $I(w)$ as a function of $w$
and $\omega_0$ for the Ising model with $\lambda_0=1.5$. Right:
Similar plot for the Kitaev model with $J_1=J_2=1$ and
$\lambda_0=2$. }\label{fig2}
\end{figure}

Let us consider the system described by $H_{\rm int}$ in its ground
state $|\psi^{\rm gnd} \rangle =\prod_{\bf k} |\psi_{\bf k}^{\rm
gnd}(0)\rangle$ at $t=0$ with $|\psi_{\bf k}^{\rm gnd}(0)\rangle=
u_{\bf k}^0 |0\rangle + v_{\bf k}^0 |1\rangle$, where
\begin{eqnarray}
u_{\bf k}^0[v_{\bf k}^0]&=& (1 +[-] (2\lambda_0-b_{\bf k})/2E_{\bf
k}(\lambda_0))^{1/2}/\sqrt{2}, \label{uveq}
\end{eqnarray}
and $|0\rangle$ and $|1\rangle$ denote the states $(1,0)\equiv
c_{1{\bf k}}^{\dagger} |{\rm vac}\rangle$ and $(0,1) \equiv c_{2{\bf
k}}^{\dagger} |{\rm vac}\rangle$ respectively for a given ${\bf k}$,
with $|{\rm vac}\rangle$ being the vacuum state for fermions. The
corresponding excited state is given by $|\psi_{\bf k}^{\rm
ex}(0)\rangle= -v_{\bf k}^0 |0\rangle + u_{\bf k}^0 |1\rangle$. The
state of the system at $t=t_f$ is given by $|\psi_{\bf
k}(t_f)\rangle= u_{\bf k} |0\rangle + v_{\bf k} |1\rangle$, where
the expressions for $u_{\bf k}$ and $v_{\bf k}$ can be obtained by
solving the Schrodinger equation $i \hbar
\partial_t |\psi_{\bf k}(t)\rangle = H_{\bf k}(t) |\psi_{\bf
k}(t)\rangle$. Thus the wavefunction overlaps $\alpha_{\bf k}=
\langle \psi_{\bf k}^{\rm gnd}(0)|\psi_{\bf k}\rangle$ and
$\gamma_{\bf k}= \langle \psi_{\bf k}^{\rm ex}(0)|\psi_{\bf
k}\rangle$ are given by
\begin{eqnarray}
\alpha_{\bf k}  &=& (u_{\bf k}^{\ast 0} u_{\bf k} + v_{\bf k}^{\ast
0} v_{\bf k}), \quad  \gamma_{\bf k} = (u_{\bf k}^{\ast 0} v_{\bf k}
- v_{\bf k}^{\ast 0} u_{\bf k}). \label{cexp}
\end{eqnarray}
We note that one can express $n_d$ and $Q$ in terms of $\gamma_{\bf
k}$ as $n_d[Q](\omega_0) = \int \frac{d^d k}{(2\pi)^d} 1[2 E_{\bf
k}] |\gamma_{\bf k}|^2$. Further, using Eqs.\ \ref{geneq2} and
\ref{feq1}, one obtains, in terms of the wavefunction overlaps,
\begin{eqnarray}
f(u,\omega_0) = - \int \frac{d^d k}{(2\pi)^d} \ln \Big[|\alpha_{\bf
k}|^2 + |\gamma_{\bf k}|^2 e^{-2 E_{\bf k}(\lambda_0) u}\Big],
\label{feq2}
\end{eqnarray}
where the integral is to be taken over the $d$-dimensional Brillouin
zone. The corresponding rate function, $I(w,\omega_0)$ can be
computed from Eq.\ \ref{feq2} using Eq.\ \ref{gerate}; in
particular, $I(0,\omega_0)$ is given by
\begin{eqnarray}
I(0,\omega_0) &=& \int \frac{d^d k}{(2\pi)^d} \ln (1 -|\gamma_{\bf
k}|^2) \ge n_d(\omega_0), \label{ieq3}
\end{eqnarray}
where we have used the expression of $n_d(\omega_0)$.

\begin{figure}[t!]
\begin{center}
\includegraphics[width=0.45\columnwidth]{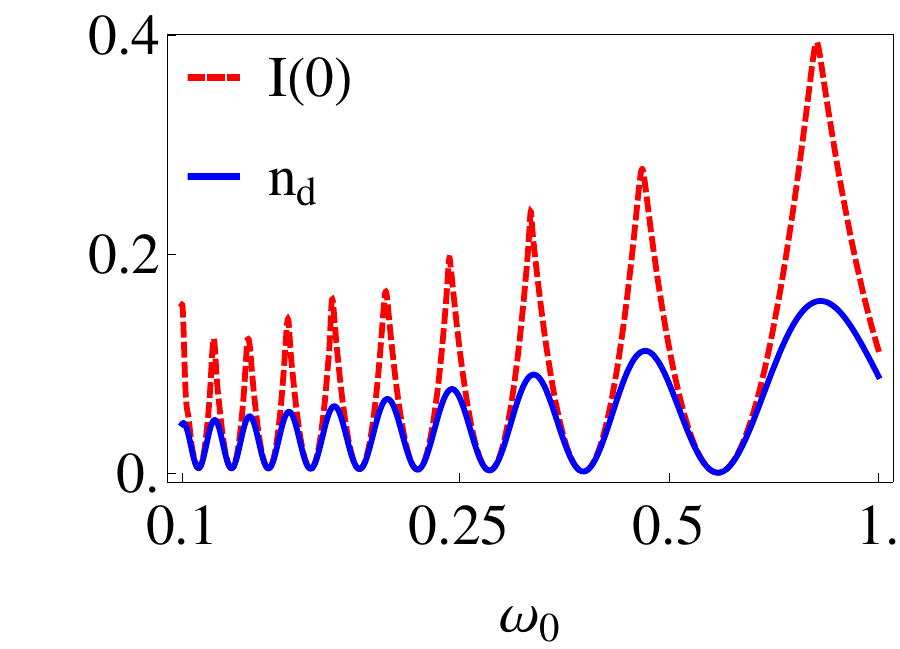}
\includegraphics[width=0.45\columnwidth]{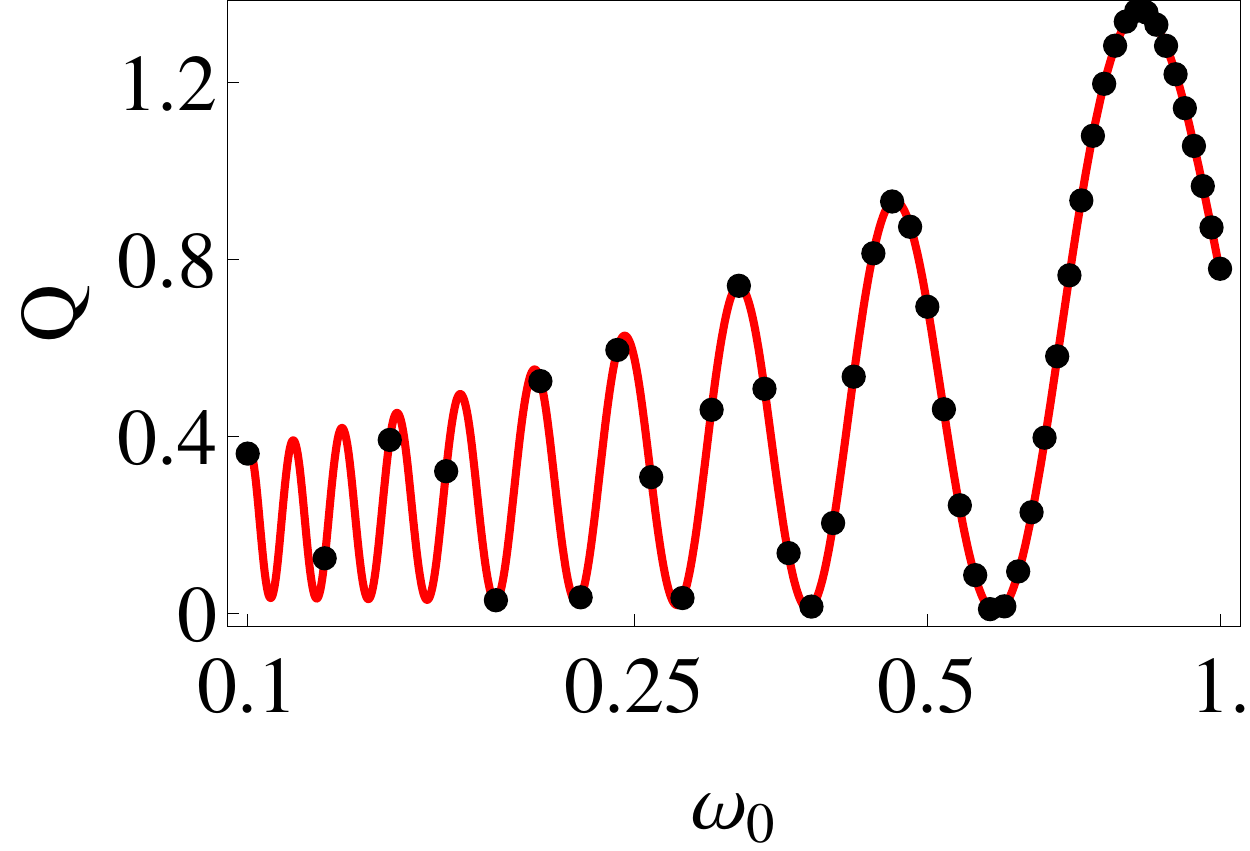}
\includegraphics[width=0.45\columnwidth]{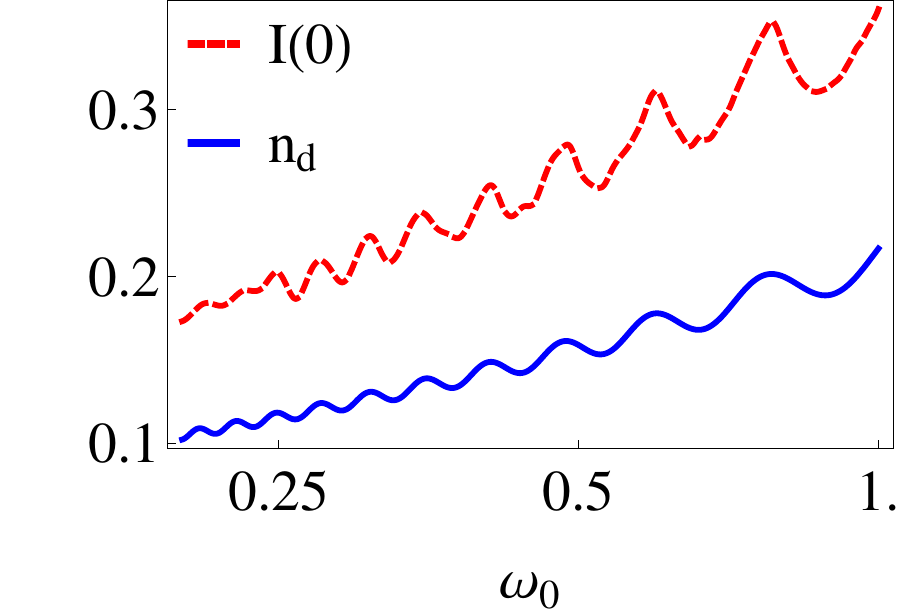}
\includegraphics[width=0.45\columnwidth]{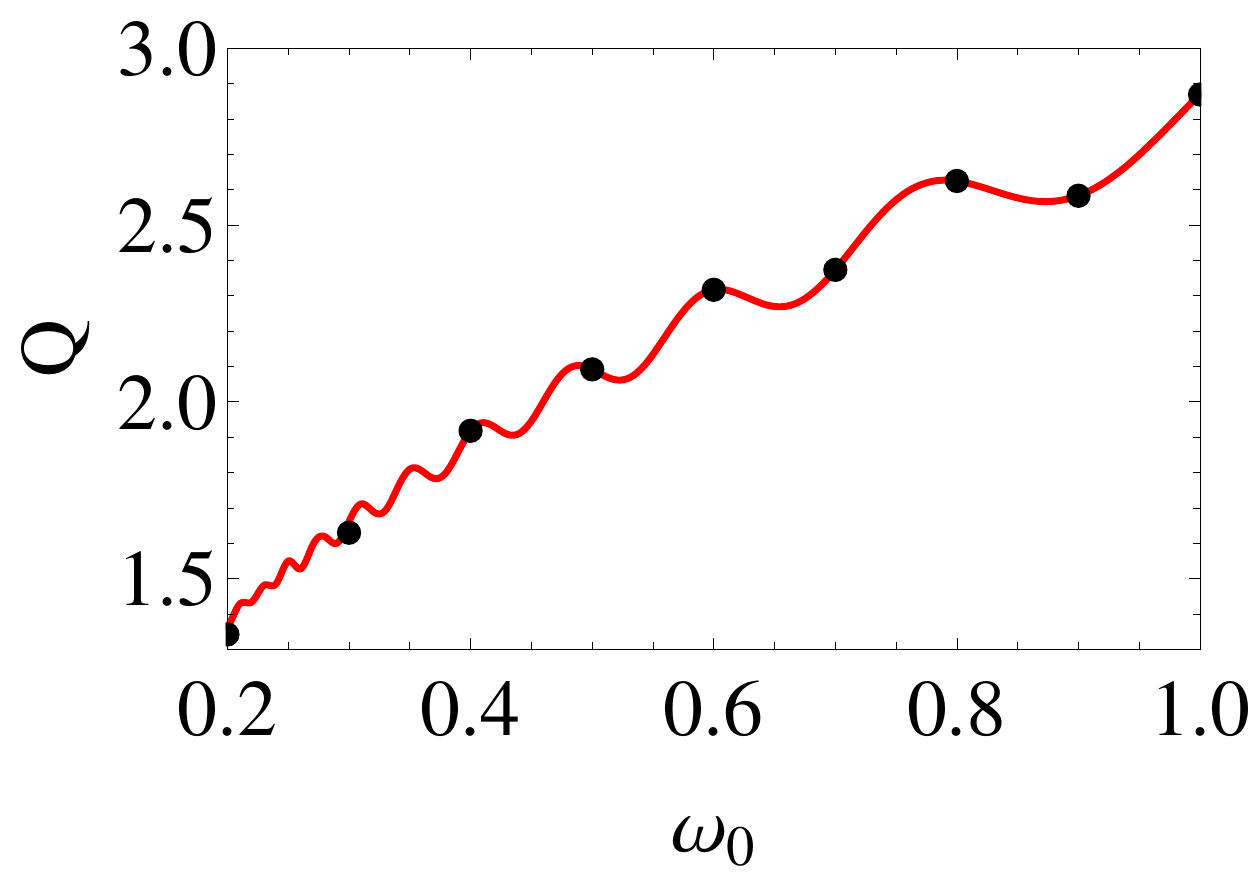}
\end{center}
\caption{(Color online) Top Left: Plot of $I(0)$ (red dashed line)
and $n_d$ (blue solid line) for $d=1$ Ising model as a function of
$\omega_0$. Top right: Plot of Q as calculated directly (red solid
line) and from $I(Q)=0$ (black dots) as a function of $\omega_0$.
Bottom panels: Similar plots for the $d=2$ Kitaev model. All
parameters are same as in Fig.\ \ref{fig2}. }\label{fig3}
\end{figure}

Eq.\ \ref{feq2} and \ref{ieq3}, along with Eq.\ \ref{gerate} reduces
the task of computing task of computing $I(w)$ to computing
$\gamma_{\bf k}$. This can be done exactly, albeit numerically, for
both the Kitaev and the Ising models as shown in Figs.\ \ref{fig2}
and \ref{fig3}. In Fig.\ \ref{fig2}, we find that for both the
models, $I(w)$ is a non-monotonic function of $\omega_0$. Further in
the left panels of Fig.\ \ref{fig3}, we confirm the validity of the
bound $I(0)\ge n_d$ for both the models via explicit computation of
$I(0,\omega_0)$ (Eq.\ \ref{ieq3}) and $n_d(\omega_0)$. We also
compute the position of zeros of $I(w,\omega_0)$ for several
representative values of $\omega_0$; a comparison of these values
with the plot of $Q(\omega_0)$ confirms the result $I(Q)=0$ (Fig.\
\ref{fig3} right panels).

Finally, we relate the non-monotonic behavior of $I(w,\omega_0)$ to
the Stuckelberg interference phenomenon. To this end, we focus on
Ising model for which one can derive an analytical expression of
$\gamma_{k}$ using the adiabatic-impulse approximation
\cite{nori1,commentstu1}. Within this approximation, excitation
production for any $k$ occurs in the impulse region near avoided
level crossings; for the rest of the evolution, the system gathers
an adiabatic phase factor. The calculation of these phase factors
and excitation probabilities during passages through the critical
point for any $k$ is identical to that for a two-level system
\cite{nori1}. The final result can be expressed as follows.
Consider, for any given $k$, $p_k$ to be the defect formation
probability during a single passage through the critical point
\cite{nori1,commentstu1}
\begin{eqnarray}
p_k &=& \exp(-2 \pi \delta_k), \quad \delta_k = g_k^2/|d
\lambda/dt|_{t=t_1}. \label{def1}
\end{eqnarray}
The expression for total defect formation probability,
$|\gamma_k|^2$, at $t=t_f$ is given in terms of these quantities as
\begin{eqnarray}
|\gamma_k|^2 &=& 4 p_k(1-p_k) \sin^2(\phi_k^{\rm st})
 \label{adimpgamma}
\end{eqnarray}
where $\phi_k^{\rm st}= \xi_{2k} +\phi_k$ is the Stuckelbeg phase
originating from the interference of parts of the system
wavefunction at ground and excited states during the second passage
through the critical point, $\xi_{2k} = \int_{t_1}^{t_2} 2E_k(t')
dt'/\hbar$ is the phase acquired during passage between the critical
point crossings at $t=t_1$ and $t_2$, and $\phi_k = -\pi/4 +
\delta_k[\ln(\delta_k)-1] + {\rm Arg} \Gamma(1-i\delta_k)$ is the
Stoke's phase \cite{nori1,commentstu1,comment3}.

\begin{figure}[t!]
\begin{center}
\includegraphics[width=0.45\columnwidth]{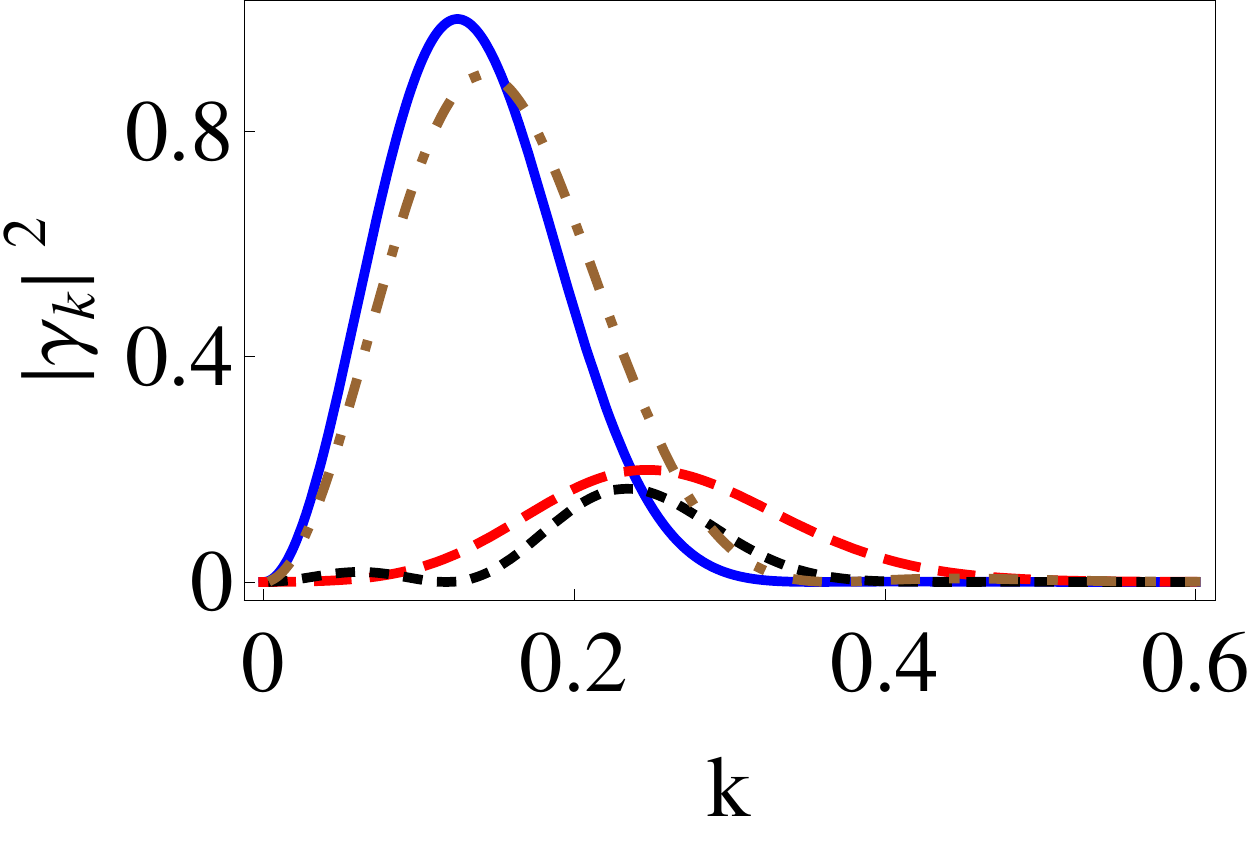}
\includegraphics[width=0.45\columnwidth]{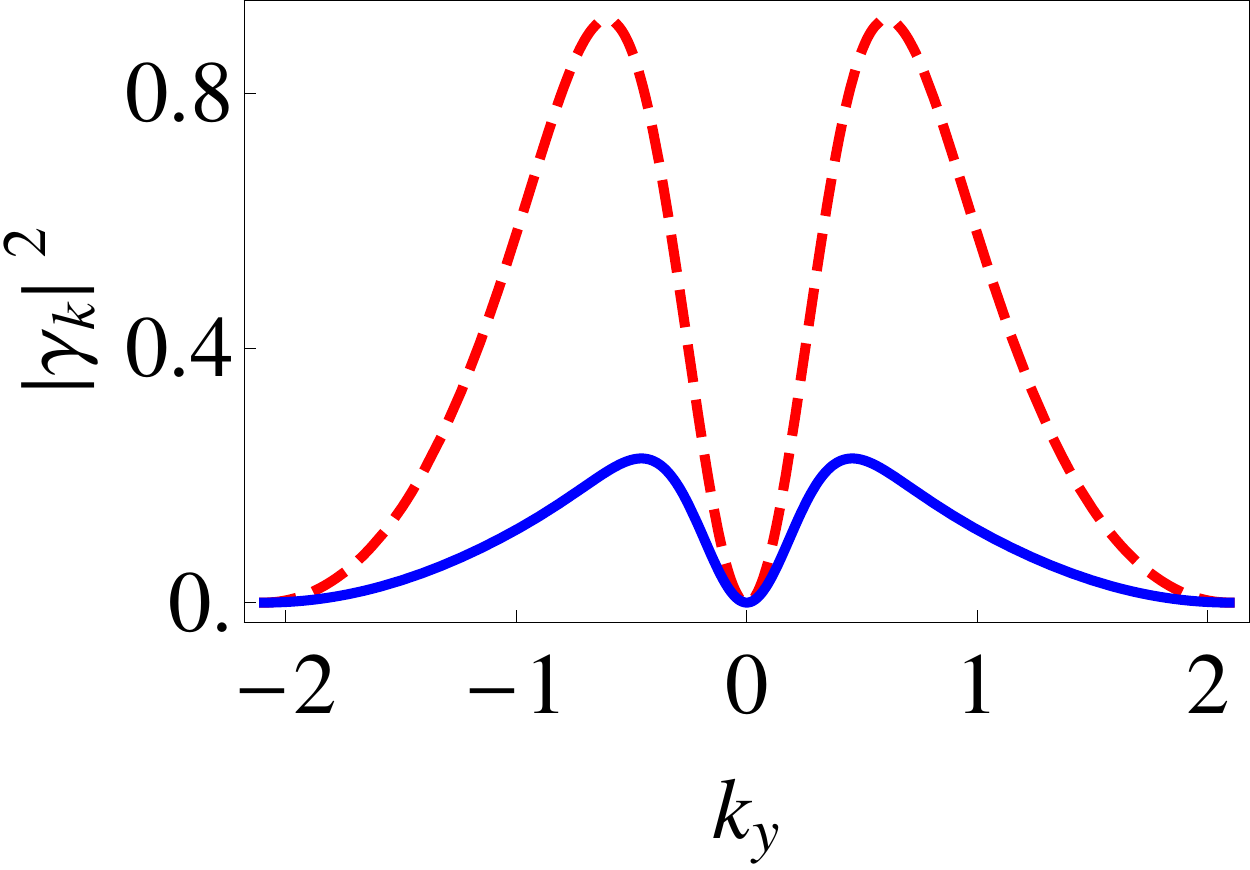}
\end{center}
\caption{(Color online) Left: Plot of $|\gamma_k|^2$ as a function
of $k$ for two representative values of $\omega_0=0.1$ [blue solid
line (numerical) and brown dash-dotted line (analytical)] and $0.12$
[red dashed line (numerical) black dashed line (analytical)]. Right:
Similar plot for $d=2$ Kitaev model with $k_x=2$ for $\omega_0=0.65$
(red dashed line) and $0.25$ (blue solid line). }\label{fig4}
\end{figure}

From Eq.\ \ref{adimpgamma}, we note that $|\gamma_k|^2 \sim 1$
provided $p_k \simeq 1/2$ and $\phi_k^{\rm st} \simeq \pi/2$ for the
same $k$; otherwise it stays small. Since $p_k$ depends on
$\omega_0$ through $\delta_k$ and $\sin^2(\phi_k^{\rm st})$ is an
oscillatory function of $\omega_0$, we expect periodic pattern of
maxima and minima for $|\gamma_k|^2$ as a function of $\omega_0$. A
plot of $|\gamma_k|^2$ vs $k$ in the left (right) panel of Fig.\
\ref{fig4} for $\omega_0=0.1$ and $0.12$ demonstrates the
above-mentioned effect for the Ising (Kitaev) model. The left panel
of Fig.\ \ref{fig4} plot also shows a qualitative match between the
analytical (Eq.\ \ref{adimpgamma}) and numerical (Eq.\ \ref{cexp})
expressions of $|\gamma_k|^2$. It is then obvious from Eqs.\
\ref{feq2} and \ref{ieq3} that such a pattern of alternate maxima
and minima will also show up in $I(w,\omega_0)$ since it depends on
momentum integral of $|\gamma_k|^2$. Thus we find that the peaks
(dips) of $I(w,\omega_0)$ arises from constructive(destructive)
interference of the ground and excited state wavefunctions; it
constitutes a manifestation of quantum interference phenomenon in
shaping the statistics of work distribution.

The experimental verification of our work would involve measuring
characteristics function $G(iu)$ leading to construction of $P(w)$
by using single qubit interferometry \cite{prop1,prop2,exp1}. Such
experiments involve spin $1/2$ systems which are implemented using
ion traps \cite{prop1} or single two-level system \cite{prop2} or
array of nuclear spins \cite{exp1,hegde1}. We suggest driving such a
system a periodic protocol with frequency $\omega_0$ for a time
$T_0= 2\pi/\omega_0$. Our theory predicts that the corresponding
$I(w)$ shall vanish at ${\bar w}=Q$ and shall satisfy $I(0)\ge n_d$.
We note that $Q$ and $n_d$ can be separately measured in these
systems \cite{commentf}. For large Ising arrays which harbors a
critical point, we also predict that $I(w)$ would be an oscillatory
function of $\omega_0$.

In conclusion, we have established that for a generic closed quantum
system subjected to a periodic drive, the rate function at the end
of a drive cycle satisfies $I(0) \ge n_d$ and $I(Q)=0$. These
relations are universal and do not depend on system or protocol
details as long the drive is periodic. We have also computed $I(w)$
for a class of integrable models where the drive takes such systems
through intermediate critical points (regions). We have shown that
$I(w)$ is an oscillatory function of $\omega_0$, linked this
behavior to the Stuckelberg interference phenomenon, and  suggested
experiments to test our theory.

\section{Supplementary Material: Work distribution for generic quantum systems}

In this section of the supplementary material, we provide the
details of the proof of the relation $I(0)\ge n_d$ for both
integrable and generic non-integrable Hamiltonians subjected to a
periodic drive.

We begin with a large class of integrable Hamiltonians for which the
many-body wavefunction can be written in product form in some
appropriate basis. This allows one to write
\begin{eqnarray}
|\psi\rangle = \prod_{j=1,N} \sum_{i=1,M} c_{i}^{j} |n_{i}^j\rangle
\label{wavefunction1}
\end{eqnarray}
Here the index $i$ runs over all possible configurations of single
particle states $|n_i\rangle$ while the index $j$ denotes the
many-body label which could either be spatial location of a system
site or its momentum. For example, for the class of integrable
models studied in the main text, $j$ indicates the momentum index
${\bf k}$ while $i$ takes value $1,2$ corresponding to the number of
single particle states for each ${\bf k}$; for instance, one may
have $c_1^{\bf k} =u_{\bf k}$ and $c_2^{\bf k}=v_{\bf k}$. The
coefficients $c_{i}^j$ satisfy $\sum_{i=1,M} |c_i^j|^2 =1$ for each
$j$. In what follows, we are going to choose a basis so that the
ground state of the system corresponds to $c_i^j=\delta_{i1}$ for
all $j$: $|\psi_{\rm gnd} \rangle= \prod_{j=1,N}  |n_1^j\rangle$.
Note that this choice do not lead to loss of generality.

The wavefunction after a drive cycle is given $|\psi_f\rangle =
\prod_{j=1,N} \sum_{i=1,M} d_{i}^{j} |n_{i}^j\rangle$. The overlap
of $|\psi_f\rangle$ with the initial ground state wavefunction is
given by
\begin{eqnarray}
|\langle \psi_f|\psi_{\rm gnd}\rangle|^2 &=& |c_0|^2 = \prod_{j=1,N}
|d_1^j|^2 = \prod_{j=1,N} (1-p_j), \label{peq1}
\end{eqnarray}
where $p_j =\sum_{i=2,M} |d_i^j|^2$ denotes the probability of
deviation from the ground state configuration for the index $j$.
This result is used in the main text. The interpretation of $p_j$
depends on the system at hand. For example, if $j$ represents
coordinates of a site of the system, it would indicate the
probability of deviation from the ground state configuration at that
site. Alternatively, if $j$ represented momentum $k$, $p_j \equiv
p_{\bf k}$ would be the probability of deviation corresponding to
the momentum mode ${\bf k}$. The existence of non-zero $p_j$
therefore indicates excitation production since it corresponds a
finite probability of the system to be in a excited state. For
integrable Hamiltonians, these excitations correspond to
quasiparticles; their total number is given by $N_0= \sum_j p_j = N
p$, where $p$ is the average probability of excitation production.
Thus the defect or excitation density of such systems are given by
$n_d = Np/L^d$,where $d$ is the dimensionality of the system and $L$
is its linear dimension. From these considerations, using Eq.\
\ref{peq1} and noting that $I(0)=-L^{-d}\ln|c_0|^2$, one finds
\begin{eqnarray}
I(0)&=& - L^{-d} \sum_j \ln(1-p_j) \ge NL^{-d} p =n_d
\end{eqnarray}
which yields the bound $I(0)\ge n_d$. Note that the equality holds
if the system comes back perfectly to the ground state after the
drive leading to $I(0)=n_d=0$; this corresponds to perfect dynamic
freezing.

We now generalize this result to generic non-integrable Hamiltonians
where the many-body wavefunction can not be written as the product
of single particle wavefunctions as in Eq.\ \ref{wavefunction1}. To
do this, we first consider the evolution operator $S = T_t (\exp[-i
\int_0^{t_f} H(t') dt'/\hbar])$ as defined in the main text which
controls the change of the many-body wavefunction during the drive.
To compute this change, we first divide the evolution time $t_f=
T_0=2\pi/\omega_0$ into ${\mathcal M}$ slices of width $\Delta t$:
$T_0= {\mathcal M} \Delta t$. A choice of small enough $\Delta t$
allows one to write
\begin{eqnarray}
S &=& \prod_{j=1, {\mathcal M}} e^{-i \int_{t_{j-1}}^{t_{j}} H(t')
dt'/\hbar} \nonumber\\
&\simeq& \prod_{j=1, {\mathcal M}}  e^{-i (H[\lambda(t_j)] +
H[\lambda(t_{j-1})])\Delta t /2\hbar} \label{sapp1}.
\end{eqnarray}
where we have used the fact that for small enough $\Delta t$ and a
generic smooth protocol, $H[\lambda(t)]$, between intervals
$t_{j-1}$ and $t_j$, can be replaced by the average of its values at
$t=t_j$ and $t_{j-1}$. This approximation is valid if $\partial_t H
\Delta t /2 \ll H$ at all times leading to
\begin{eqnarray}
&& 2 \pi V_j \ll {\mathcal M} H_j \omega_0 \label{cond1}
\end{eqnarray}
where $V_j = \partial_t H[\lambda(t)]|_{t=t_j}/2$ and we have used
$\Delta t = 2\pi/({\mathcal M} \omega_0)$. In this context, we note
a couple of points. First the condition given in Eq.\ \ref{cond1}
can be satisfied for any smooth protocol by choosing a large enough
${\mathcal M}$ and second, the adiabatic limit for the drive can be
obtained in this formalism by choosing $V_j \to 0$ for all $j$. In
what follows we denote the eigenstates and eigenenergies of $H_j$ by
$|n_j\rangle$ and $E_{n_j}$ respectively with the convention that
the ground state is given by $|0_j\rangle$.

Let us now consider that the wavefunction $|\psi_j\rangle $ at
$t=t_j$ can be expressed in the eigenbasis of $H_j$ as $|\psi_j
\rangle= \sum_{n_j} a_{n_j}^j |n_j\rangle$, where $a_{n_j}^j$
denotes the overlap of $|\psi_j\rangle$ with $|n_j\rangle$. Within
each interval $\Delta t$, the evolution of $|\psi\rangle$ is
determined by the time-independent Schrodinger equation with
effective Hamiltonian $H_j + \Delta t V_j$ and can be estimated by
time-independent perturbation theory since $\Delta t V_j \ll H_j$
(Eq.\ \ref{cond1}). Consequently, one has
\begin{eqnarray}
|n_{j+1}\rangle &=& (1-p_{n_j}^j/2)|n_j\rangle + \sum_{m_j \ne n_j}
\alpha_{m_j n_j}^j |m_j\rangle \label{wavrel} \\
\alpha_{m_j n_j}^j&=& \frac{\langle m_j |\Delta t V_j|n_j
\rangle}{E_{m_j}-E_{n_j}} +{\rm O}(\Delta t^2), \,\, p_{n_j}^j =
\sum_{m_j \ne n_j } |\alpha_{m_j n_j}|^2 \nonumber\
\end{eqnarray}
We note that the formal requirement for the convergence of the
second order perturbation theory used here is $|\alpha_{m_j n_j}^j
|\ll 1$. This leads to an estimate of minimum value of ${\mathcal
M}$. To see this, we first consider a typical many-body system where
the minimum energy spacing between two states in the Hilbert space
goes as $\exp[-N]$ where $N$ denotes the number of sites in the
system. So one needs ${\mathcal M} \ge \exp[N]$ for a generic
protocol for such a perturbation theory to converge. In this
context, we note two points. First, for slow dynamics, the system
may only traverse low-lying excited states where the energy gap, in
the presence of a critical point, goes as $\Delta E \sim N^{-z}$,
where $z$ is the dynamical critical exponent. For such protocols, it
suffices to use ${\mathcal M} \sim N^z$. Second, our formal argument
leading to the final result is completely independent of the precise
choice of ${\mathcal M}$ as long as it is large enough for the
perturbation theory to converge and, consequently, for $\alpha_{m_j
n_j}^j$ and $p_{n_j}^j$ to remain small.

Since the wavefunction at $t=t_{j+1}$, $|\psi_{j+1}\rangle$, can be
expressed as $|\psi_{j+1}\rangle = \sum_{n_{j+1}}  a_{n_{j+1}}^{j+1}
|n_{j+1}\rangle$, one can obtain, using Eq.\ \ref{wavrel}, a
recursion relation for $a^j_{n_j}$:
\begin{eqnarray}
a_{n_{j+1}}^{j+1} &=& (1-p_{n_j}^j/2) a_{n_j}^j + \sum_{m_j \ne n_j}
\alpha_{m_j n_j}^j a_{m_j}^j.  \label{coeffrel1}
\end{eqnarray}
Using the fact that the system starts in the ground state at $t=0$
so that $a_{0_0}^0=1$ and $a_{m_0}^0=0$ for $m_0 \ne 0$ , we can
expand Eq.\ \ref{coeffrel1} to obtain
\begin{eqnarray}
a_{0_1}^1 &=& (1-p_0/2), \quad a_{m_1}^1 = \alpha_{m_0 0_0}^0
\nonumber\\
a_{0_2}^2 &=& (1-p_0/2)(1-p_1/2) + \sum_{m_1} \alpha_{m_0 0_0}^0
\alpha_{m_1 0_1}^1 \nonumber\\
a_{m_2}^2 &=& (1-p_{m_1}^{1}/2) \alpha_{m_0 0_0}^0 + (1-p_0/2)
\alpha_{m_1 0_1}^1 \nonumber\\
&& + \sum_{n_2 \ne m_2 \ne 0} \alpha_{n_1 m_1}^1 \alpha_{n_0 m_0}^0
\label{receq2}
\end{eqnarray}
where we have used $p_{0_j}^j \equiv p_j$. One can similarly obtain
expressions for $a_{0_j}^j$ for $j>2$ which gets increasingly
cumbersome with increasing $j$.

The wavefunction overlap after ${\mathcal M}$ such step is given by
$c_0 = \langle \psi_{\mathcal M}|0\rangle= a_{0_{\mathcal
M}}^{\mathcal M}$. This is a consequence of the periodicity of the
drive which leads to $|0_{\mathcal M}\rangle= |0_0\rangle \equiv
|0\rangle$. Using Eq.\ \ref{coeffrel1} and \ref{receq2}, we obtain,
after some algebra,
\begin{eqnarray}
c_0 &=& \prod_{j=0}^{{\mathcal M}-1} (1-p_j/2) +
\sum_{j=1}^{{\mathcal M}-1} \sum_{n=1}^{{\mathcal M}-j}
\alpha_{m_{j-n} 0_{j-n}}^{j-n} \alpha_{m_{j} 0_{j}}^{j},
\label{coeq1}
\end{eqnarray}
where we have ignored terms ${\rm O}(\Delta t^3)$ or higher. We note
that $c_0$ is a complex valued polynomial function of ${\rm
O}({\mathcal M})$ with ${\mathcal M}$ zeroes; consequently, it can
always be written in the form $c_0 = \prod_{j=0,{\mathcal M}-1}
(1-p_{1j}/2)$.

To find an expression for $p_{1j}$, we first note that each term in
the second sum in the right side of Eq.\ \ref{coeq1} is bounded. To
see this we note that from Eq.\ \ref{wavrel}, we have $\sum_{m_j \ne
0_j} |\alpha_{m_j 0_j}^j|^2 =p_j$. Thus one can write
\begin{eqnarray}
\sum_{m_j \ne 0_j} \alpha_{m_j 0_j}^j \alpha_{m_{j-n} 0_{j-n}}^{j-n}
= b_j^n e^{i\phi_j^n} {\rm Max}[p_j, p_{j-n}] \label{ceq}
\end{eqnarray}
where $0\le b_j^n \le 1$. The precise value of $b_j^n$, $\phi_j^n$
etc depends on system parameters as well as the details of the drive
protocol. However, since $p_j \sim {\rm O}(\Delta t^2)$ for any $j$,
one can see that $\sum_n \sum_{m_j \ne 0_j} \alpha_{m_j 0_j}^j
\alpha_{m_{j-n} 0_{j-n}}^{j+n}$ is at most ${\rm O}(\Delta t)$. Thus
to ${\rm O}(\Delta t^2)$, one obtains
\begin{eqnarray}
p_{1j} &\simeq& p_j - 2\sum_{n=1}^{{\mathcal M}-j}\sum_{m_j \ne 0_j}
\alpha_{m_j 0_j}^j \alpha_{m_{j-n} 0_{j-n}}^{j-n} \nonumber\\
|c_0|^2 &\simeq& \prod_{j=0,{\mathcal M}-1} (1-p'_j), \,\,  p'_j =
{\rm Re}[p_{1j}] - \frac{|p_{1j}|^2}{4} \label{p1eq}
\end{eqnarray}
which is the result used in the main text.

Next, we relate $p'_j$ to the excitation density. To do this, we
note that for any evolution step $j$, one can use Eqs.\
\ref{coeffrel1} and \ref{receq2} to show that
$a_{0_{j+1}}^{j+1}-a_{0_j}^j = -p_{1j}$ to ${\rm O}(\Delta t^2)$.
This leads to
\begin{eqnarray}
|a_{0_{j+1}}^{j+1}/a_{0_j}^j|^2 = (1-p'_j) \label{probeq}
\end{eqnarray}
Since $a_{0_j}^j$ denote the probability amplitude of the system to
be in the ground state $|0_{j}\rangle$, we find that $p'_j$ denotes
the change in the system wavefunction overlap with the instantaneous
ground state during the $j^{\rm th}$ evolution step. Since we start
from the ground state configuration at $t=0$, the average
probability of such deviation during the drive cycle is $p=
\sum_{j=0,{\mathcal M}-1} p'_j/{\mathcal M}$; thus the total number
of excitations in the system at the end of the drive is given by
$N_0={\mathcal M} p$. In this context, we note a few points. First,
since $|c_0| \le 1$ for any unitary evolution, $\sum_{j}p'_j \ge 0$;
however an individual $p'_j$ can either be positive or negative
since the system wavefunction overlap with the instantaneous ground
state may increase or decrease during the drive at the $j^{\rm th}$
step. Second, in a generic non-integrable system, the nature of
these excitations need not be quasiparticles-like; they could be,
for example, non-local objects such as vortices or extended defects.
Third, although $p'_j$ depends on the choice of ${\mathcal M}$,
$\sum_j p'_j$ and consequently $N_0$ is independent of such choice
as long as ${\mathcal M}$ satisfies the criteria of convergence of
perturbation theory discussed earlier. As ${\mathcal M}$ is
increased beyond its minimum allowed value, $\sum_j p'_j$ is
expected to converge to a fixed value independent of ${\mathcal M}$.
Thus we find $ n_d= N_0/L^d$. Using these results we finally get
\begin{eqnarray}
I(0) = -L^{-d} \sum_{j=1}^{\mathcal M} \ln (1-p'_j) \ge L^{-d} N_0
=n_d
\end{eqnarray}
which is used in the main text. Note that for slow dynamics in a
critical system with $z=1$, one may choose ${\mathcal M}=N$ as
discussed earlier leading to $\ln|c_0| \sim N$ which is known in the
standard literature \cite{intref}.

Finally, we consider generalization of our arguments for system in a
thermal/non-equilibrium diagonal ensemble at $t=0$ such that the
probability of occupation of the state $|\alpha\rangle$ is $\rho_{0
\alpha}$. Such a quantum system in an ensemble characterized by
$\rho_{0 \alpha}$, in contrast to the zero temperature case where
one starts the time evolution starts from the initial ground state
$|0\rangle$ at $t=0$, has a finite weight in all states
$|\alpha\rangle$. Thus the value of $w$ can take both positive and
negative values $E_{\alpha}-E_{\beta}$ which leads to
\begin{eqnarray}
P(w) &=& \sum_{\alpha \beta} P_{\alpha \to \beta} \delta
(w-E_{\beta} + E_{\alpha}) \nonumber\\
P_{\alpha \to \beta} &=& P_{\alpha}^0 P_{\beta \| \alpha} =
\rho_{0\alpha} |\Lambda_{\alpha \beta}|^2, \label{workdistfi}
\end{eqnarray}
where $P_{\alpha}^0=\rho_{0\alpha}$ is the probability that the
system starts out in the state $|\alpha\rangle$ and $P_{\beta \|
\alpha}$ is the conditional probability that it ends up in the state
$|\beta\rangle$ after a drive cycle having started from the state
$|\alpha \rangle$. For a thermal ensemble, for example, $\rho_{0
\alpha}= \exp(-\beta E_{\alpha})/Z$, where $E_{\alpha}$ are the
eigenenergies of the system satisfying $H[\lambda_0] |\alpha\rangle=
E_{\alpha} |\alpha\rangle$, $\beta= (k_BT)^{-1}$ is the inverse
temperature, $k_B$ is the Boltzman constant, and $Z =\sum_{\alpha}
\exp(-\beta E_{\alpha})$ is the partition function. The precise form
of the occupation probability $\rho_{0 \alpha}$ shall be irrelevant
in our derivation. We shall, however, use the relation
$\sum_{\alpha} \rho_{0 \alpha}=1$ which is generically true for any
ensemble. Also, in what follows, we are going to concentrate on
integrable systems for which $E_{\alpha} = \sum_j E^j n^j_{\alpha}$,
where $E^j$ are the single particle excitation energies
corresponding to the mode $j$ and $n_{\alpha}^j$ are the occupation
number of the states with energy $E^j$; consequently $\rho_{0
\alpha} = \prod_{j=1,N} \rho_{0 \alpha}^j$, where $\rho_{0
\alpha}^j= \exp(-\beta E^j n^j_{\alpha} )/Z$.

Now let us imagine driving this system with periodic protocol for a
time $T_0$. At the end of this drive cycle, one can write
$S|\alpha\rangle= \sum_{\beta} \Lambda_{\alpha \beta}
|\beta\rangle$, where $\Lambda_{\alpha \beta}$ are the wavefunction
overlap between the states $S|\alpha\rangle$ and $|\beta\rangle$. In
this notation, $\Lambda_{1 \beta} =c_{\beta}$ which is used in the
main text and for any $|\alpha \rangle$, $\sum_{\beta}
|\Lambda_{\alpha \beta}|^2 =1$. After the drive, the conditional
probability of the system to be in a state $|\beta\rangle$, provided
it started from a state $|\alpha\rangle$ is given by $\langle \beta|
S |\alpha\rangle|^2 = |\Lambda_{\alpha \beta}|^2$. For integrable
models, such a wavefunction overlap can be written as
$\Lambda_{\alpha \beta} = \prod_{j=1,N} \Lambda_{\alpha \beta}^j$,
where $\Lambda_{\alpha \beta}^j$ correspond to the overlap of the
$j^{\rm th}$ mode. Thus, starting from ensemble characterized by
$\rho_{0 \alpha}$, the probability of the system to return to the
same configuration after a drive cycle, $|c_0|^2$, is given by
\begin{eqnarray}
|c_0|^2 &=&  \sum_{\alpha} \rho_{0 \alpha} |\Lambda_{\alpha
\alpha}|^2  = \prod_{j=1,N} \Big [1-\sum_{\beta \ne \alpha} \rho_{0
\alpha}^j |\Lambda_{\alpha \beta}^j|^2 \Big]. \nonumber\
\end{eqnarray}
This leads to $p_j =\sum_{\beta \ne \alpha} \rho_{0 \alpha}^j
|\Lambda_{\alpha \beta}^j|^2$. Note that if the system started from
a quantum ground state $|0\rangle$ at $T=0$ (as in the main text),
$\rho_{0 \alpha} = \delta_{\alpha 1}$ and $|\Lambda_{11}^j|^2 =
|d_{1j}|^2$, so that one obtains $|c_0|^2=\prod_{j=1,N}
(1-\sum_{i\ne 1} |d_{ij}|^2)$ in this limit. Next, we note that any
deviation from the starting configuration of the system indicates
change in quasiparticle number due to the drive; this is understood
by appealing to the fact that such deviation represents energy
absorption/emission by the system during the drive which occurs via
quasiparticle production in integrable systems. In contrast to the
system at $T=0$, the change in quasiparticle number can be negative
in this case; however, the absolute value of this change is still
given by $Np_0 = \sum_j p_j$. Thus one can define a defect density
$n_d= N p_0/L^d$, where $N$ is the total number of degrees of
freedom in the system and $L$ is its linear dimension as defined in
the main text. This leads to the bound $I(0) \ne n_d$ for integrable
systems in a thermal or diagonal non-equilibrium ensemble. We note
that our definition of $n_d$ here matches with that in standard
literature \cite{rev1}

\section{Supplementary Material: Equilibrium Phase diagram of the Ising and the Kitaev
model}

In this section, we provide a brief introduction to the equilibrium
phase diagram of the Ising and the Kitaev models. These are already
discussed at length in standard literature \cite{isingbook,
kitaevref}. Here we briefly outline the main points which would be
useful in our discussion of the periodic dynamics of these models in
the main text.

The 1D Ising model in the presence of a transverse field has a
Hamiltonian given in Eq.\ 11 of the main text
\begin{eqnarray}
H_{\rm Ising} &=& -J \sum_{\langle ij\rangle} S_i^z S_j^z -h \sum_i
S_i^x, \label{ising1}
\end{eqnarray}
where $J$ is the nearest-neighbor coupling between the spins and $h$
is the transverse field and $S^{\alpha}$ denotes half-integer spins.
The exact solution of this model leading to its ground state for any
value of dimensionless transverse field $g=h/J$ can be obtained
mapping the spins to fermions via a Jordan-Wigner transformation
\cite{isingbook}
\begin{eqnarray}
\sigma_i^x = (1-2c_i^{\dagger} c_i), \quad \sigma_i^z =
\left(\prod_{j<i} \sigma_j^x \right) (c_i+c_i^{\dagger})
\label{jwtrans}
\end{eqnarray}
Substituting Eq.\ \ref{jwtrans} in Eq.\ \ref{ising1}, one obtain the
Ising Hamiltonian in terms of the free fermions: $H= \sum_{k>0}
\psi_k^{\dagger} H_k \psi_k$, where $\psi_k \equiv (c_{1k},c_{2k})^T
= (c_k, c_{-k}^{\dagger})^T$, $c_k = \sum_j \exp[ikj] c_j$, $H_k$ is
given by
\begin{eqnarray}
H_k &=& [g-\cos(k)] \tau_z + \sin(k) \tau_x, \label{fermham}
\end{eqnarray}
and we have set the lattice spacing to unity. Since Eq.\
\ref{fermham} represents a free fermion Hamiltonian, its eigenvalues
can be readily found out by diagonalizing $H_k$ and are given by
$E_k^{\pm} = \pm \sqrt{1+g^2-2g \cos(k)}$. The phases of the model
can be understood in this language as follows. For $g \gg 1$, the
ground state corresponds to the Fermion vacuum which corresponds to
(Eq.\ \ref{jwtrans}) $S_x^i=\hbar/2$ on every lattice site; this is
the paramagnetic phase of the model. For $g \ll 1$, the ground state
is an eigenstate of $\sigma_z$ which in the fermionic language is
linear superposition of $c_k$ and $c_{-k}^{\dagger}$; this
corresponds to the ferromagnetic state \cite{isingbook}. In between
at $g=1$, there is a quantum phase transition between these two
states which can be seen, in the fermionic language, by noticing
that the energy gap $\Delta_k = (E_k^+-E_k^-)$ vanishes at $g=1$ and
$k=0$ signifying a gapless spectrum at this point.

The Kitaev model, describing half integer spins on a 2D honeycomb
lattice, has the Hamiltonian \cite{kitaevref}
\begin{eqnarray}
H' &=& \sum_{j+\ell ={\rm even}}  J_{1} S_{j,\ell}^{x}
S_{j+1,\ell}^{x} + J_{2} S_{j,\ell}^{y} S_{j-1,\ell}^{y}
\nonumber\\
&& +J_3 S_{j,\ell}^{z} S_{j,\ell+1}^{z}, \label{kiteq1}
\end{eqnarray}
where $J_{1,2,3}$ denote nearest neighbor coupling between the
spins, $(j,\ell)$ describe 2D lattice coordinates, and
$S_{j,\ell}^{\alpha}$ for $\alpha=x,y,z$ denotes half-integer spins.
Note that the basic difference of this model with, for example, the
anisotropic Heisenberg model, is that the coupling between a pair of
spins on the neighboring sites of the lattice in the $\alpha^{\rm
th}$ direction involves only $S_{j \ell}^{\alpha}$. This feature
makes the model solvable; in fact, the Kitaev model constitutes one
of the very few examples of a solvable interacting quantum model in
$d>1$.

The steps to solving of $H'$ proceeds in the same manner as those of
the 1D Ising chain. First, one designs a Jordan Wigner
transformation relating the spins $S_{j,\ell}^{\alpha}$ to Majorana
fermion operators \cite{kitaevref}
\begin{eqnarray}
a_{j \ell} [a'_{j \ell}] &=& \left(\prod_{i=-\infty}^{j-1}
\sigma_{i\ell}^z \right) \sigma_{j\ell}^{y} [\sigma_{j\ell}^{x}]
\quad {\rm \, for \, even}\, j +\ell
\nonumber\\
b_{j \ell} [b'_{j \ell}]&=& \left(\prod_{i=-\infty}^{j-1}
\sigma_{i\ell}^z \right) \sigma_{j\ell}^{x} [\sigma_{j\ell}^{y}]
\quad {\rm \, for \, odd} \,j +\ell  \label{2djw}
\end{eqnarray}
Substituting Eq.\ \ref{2djw} in Eq.\ \ref{kiteq1}, one obtains a
fermionic Hamiltonian given by
\begin{eqnarray}
H_0 &=& i \sum_{\vec n} J_1 a_{\vec n} b_{\vec n +\vec M_1} + J_2
a_{\vec n} b_{\vec n +\vec M_2} + J_3 D_{\vec n} a_{\vec n} b_{\vec
n} \label{kitferm}
\end{eqnarray}
where $\vec n = \sqrt{3} n_1 \hat i + n_2 \sqrt{3}\hat i + \hat
j)/2$ denotes midpoints of the vertical bonds, $n_1$ and $n_2$ run
over all integers, $\hat i$ and $\hat j$ denote unit vectors along
$x$ and $y$ directions, ${\bf M}_{1,2}= (\sqrt{3}/2,+(-)3/2)$ denote
the spanning vectors of the reciprocal lattice of the model and we
have set the lattice spacing to unity. Here $D_{\vec n} = a'_{\vec
n} b'_{\vec n}$ and can take value $\pm 1$ for each $n$. The key
point that makes the Kitaev model exactly solvable is that $D_{\vec
n}$ commutes with $H_0$; thus one can find the energy spectrum
corresponding to a set of given values of $D_{\vec n}$. As shown in
Ref.\ \onlinecite{kitaevref}, the ground state corresponds to
$D_{\vec n}=1$ on all links. Further, since $D_{\vec n}$ commutes
with $H_0$ its values does not change on making $H$ time dependent
via $J_3 \to J_3(t)$. Thus, for our purpose, it would be sufficient
to set $D_{\vec n}=1$. A Fourier transform of $H_0$ then leads to $
H_0 = \sum_{\vec k} \psi_{\vec k}^{\dagger} H_{\vec k} \psi_{\vec
k}$, where $\psi_{\vec k} \equiv (c_{1\vec k}, c_{2\vec k})^T=
(a_{\vec k}, b_{\vec k})^T$, $a_{\vec k}$ and $b_{\vec k}$ denote
complex fermions so that the sum over $\vec k$ is to be taken over
half of the 2D Brillouin zone, and $H_{\vec k}$ is given by
\begin{eqnarray}
H_{\vec k} = (\lambda_0 -b_{\vec k}) \tau_3 + g_{\vec k} \tau_1
\label{kitmoment}
\end{eqnarray}
where $\lambda_0=J_3$, $b_{\vec k}= (J_1 \cos({\vec k}\cdot {\vec
M_1}) + J_2 \cos({\vec k}\cdot {\vec M_2}))$, and $g_{\vec k}= (J_1
\sin({\vec k}\cdot {\vec M_1}) - J_2 \sin({\vec k}\cdot {\vec
M_2}))$.

The energy spectrum of the model can be obtained by diagonalization
of $H_{\vec k}$ and is given by $E_{\vec k}^{\pm}= \pm \sqrt{
(\lambda_0-b_{\vec k})^2 +g_{\vec k}^2}$. The phases of the model
consists of a gapless phase where $\Delta_{\vec k} = (E_{\vec
k}^+-E_{\vec k}^-) =0$; this occurs at $\vec k= \vec k_0$ satisfying
$\lambda_3 = b_{\vec k_0} $ and $g_{\vec k_0}=0$. The gapless phase
occurs when $(J_1+J_2) \ge J_3 \ge |J_1-J_2|$. For all other values
of the coupling, the model exhibits gapped ground states as shown in
Fig.\ \ref{fig1} of the main text. The properties of these phases
are discussed in details in Ref.\ \onlinecite {kitaevref}. The main
notable point for us about them is that they do not correspond to
any local spin order and constitute examples of gapped and gapless
$Z_2$ spin liquids.

\section{Supplementary Material: Stuckelberg Interference Phenomenon}

The Stuckelberg interference phenomenon usually refers to quantum
mechanical interference between amplitudes of occupation of energy
levels of a two-level system upon multiple passages through its
avoided level crossing \cite{nori1}. Such passages usually occur for
periodic drive protocols; in what follows we are going to consider a
protocol which leads to two passages of the system through such an
avoided level crossing. Under such a protocol, the system, starting
from the ground state (lower level) at $t=t_0$ when the drive
commences, transfer part of its amplitude to the upper level upon
first passage through the avoided level crossing. The interference
between these two amplitudes (probability amplitudes of occupation
of the upper and lower levels) on the second passage leads to an
oscillatory behavior of the probability of defect formation: $P_1=
4P(1-P) \sin^2\phi^{\rm st}$, where $P$ denotes the probability of
the system to be in the excited state after the first passage, and
$\phi^{\rm st}$ is the Stuckelbreg phase which depends on the drive
frequency and amplitude. Usually, this phase has two parts; the
first, $\xi$ originates from the relative phase picked up by the
system in between its passages through the avoided level crossing,
while the second, $\phi$, known as the Stokes's phase, is picked up
during the second passage through the avoided level crossing:
$\phi^{\rm st}= \phi+ \xi$ \cite{nori1}. An analytical description
of this phenomenon leading to the expression of $P_1$ presented
above is not exact; it requires the so called adiabatic-impulse
approximation and has been detailed in Ref.\ \cite{nori1}. Within
this approximation, the transition between the two levels occur only
near the avoided level crossing; the rest of the evolution of the
system is assumed to be near-adiabatic. This approximation captures
the essential qualitative aspects of this phenomenon and is also
quantitatively accurate for large amplitudes and/or low frequency
drive protocols. Also, for a class of drives which corresponds to a
series of rectangular periodic pulses, this approximation becomes
exact.

Next, we briefly outline the derivation of $p_k$ and $\phi_k^{\rm
st}$ which is used in Eqs.\ 17 and 18 of the main text. To this end,
we note that the Ising model, discussed in the main text, reduces to
a series of two-level systems with each $k$ and thus the results of
Ref.\ \onlinecite{nori1} can be directly adapted. We begin by
assuming that the Ising model is in its ground state which, in the
fermionic representation, corresponds to $|\psi_k^0\rangle = u_k^0
|0\rangle + v_k^0 |1\rangle$ for each $k$, where $|0\rangle \equiv
(1,0) \equiv c_{1k}^{\dagger} |{\rm vac}\rangle$, $|1\rangle \equiv
(0,1) \equiv c_{2k}^{\dagger} |{\rm vac}\rangle$, and $|{\rm
vac}\rangle$ denotes fermionic vacuum. The corresponding excited
state is given by $|\psi_k^1\rangle = -v_k^0 |0\rangle + u_k^0
|1\rangle$. The Hamiltonian of the system is given by Eq.\ 10 of the
main text. We now consider evolution of the system under a
oscillatory drive protocol given by $\lambda(t)=
\lambda_0[1+\cos(\omega_0 t)]$. We note that under this protocol the
system passes through the critical points at $\omega_0 t_1=
\arccos(b_{k_0}/\lambda_0-1)$ and $t_2= 2\pi/\omega_0 -t_1$, where
$k_0$ is the critical wave-vector at which the gap closes. For the
$d=1$ Ising model in a transverse field, $k_0=0$.

The wavefunction of the system at a time $t$ can be written as
\begin{eqnarray}
|\psi(t)\rangle = b_{1k}(t) |\psi_k^0 \rangle + b_{2k}(t) |\psi_k^1
\rangle \label{wav1}
\end{eqnarray}
where $b_k =(b_{1k}(t),b_{2k}(t))$ can be found out by solving the
Schrodinger equation with the initial condition $b_{k}(t=t_0)=(1,0)$
\cite{nori1}. The exact analytical solution for $b_k(t)$ does not
exist for the protocol studied here; however, an approximate
solution within the adiabatic-impulse approximation can be found out
following the method outlined in Ref.\ \onlinecite{nori1}. Within
this approximation, the evolution of $b_k$ gets contribution from
two parts. The first is a phase gathered during adiabatic evolution
away from the avoided level crossing. This is described by $U_k$
given by
\begin{eqnarray}
U_k(t,t') &=&  I e^{i \tau_3 \xi_k(t,t')}, \quad
\xi_k(t,t') = \frac{1}{2} \int_{t'}^{t} E(t'') dt'', \nonumber\\
\label{uexp}
\end{eqnarray}
where $E_k(t)$ denotes the instantaneous eigenenergy corresponding
to the wavevector $k$ and $\tau_{i}$, for $i=1,2,3$, are the Pauli
matrices with the convention that $(1,0)^T$ is the eigenfunction of
$\tau_3$. The second contribution to $b_k$ comes during passage of
the system through the critical point where the dynamics is no
longer adiabatic and defect formation occurs. Around this point, it
is possible to linearize the drive term: $\lambda(t) \simeq (t-t_a)
d\lambda(t)/dt|_{t=t_a}$, where $t_a$ denotes the time when the
system passes through the critical point; consequently, the dynamics
reduces to an effective LZ problem for each $k$ and can be exactly
solved. The contribution to $b_k(t)$ due to passage through such
avoided level crossing  is described by the matrix $N_k$ given by
\cite{nori1}
\begin{eqnarray}
N_k &=& \sqrt{1-p_k}e^{i\tau_3 \phi_k} -i\tau_2 \sqrt{p_k},
\nonumber\\
p_k &=& \exp(-2 \pi \delta_k), \quad \delta_k = g_k^2/|d \lambda/dt|_{t=t_1}  \nonumber\\
\phi_k &=& -\pi/4 + \delta_k[\ln(\delta_k)-1] + {\rm Arg}
\Gamma(1-i\delta_k), \label{nexp}
\end{eqnarray}
where $p_k$ is the probability of excitation production upon a
single passage through a critical point and $\phi_k$ is the Stoke's
phase for a given $k$. In terms of these matrices, one can write
\cite{nori1}
\begin{eqnarray}
b_k(t_f) &=& U_k(t_f,t_2)N_{2k} U_k(t_2,t_1) N_{1k} U_k(t_1,t_i)
b_k(t_i) \nonumber\\ \label{bexp1}
\end{eqnarray}
A few lines of straightforward algebra then yields
\begin{eqnarray}
b_{1k}(t_f) &=& [(1-p_k)e^{-i \xi_{+k}} -p_k e^{-i \xi_{-k}}]e^{-i\xi_k(t_1,t_i)} \nonumber\\
b_{2k}(t_f) &=& \sqrt{p_k(1-p_k)} e^{i \phi_k}
(e^{-i\xi_{+k}} + e^{-i \xi_{-k}}) \nonumber\\
\xi_{+k} &=& \xi_k(t_2,t_1) + \xi(t_1+2\pi/\omega_0,t_2) +2 \phi_{k} \nonumber\\
\xi_{-k} &=& \xi_k(t_2,t_1) - \xi(t_1+2\pi/\omega_0,t_2)
\label{evol1}
\end{eqnarray}
where we have assumed that the system starts its evolution at
$t=t_i$, ends it at $t=t_f$, and crosses the critical point at
$t=t_1$ and $t_2$. Within the adiabatic-impulse approximation, $u_k
\equiv u_k(t_f)$ and $v_k \equiv v_k(t_f)$ is therefore given by
\begin{eqnarray}
u_k(t_f) &=& u_k^0 b_{1k}(t_f) -v_k^0 b{2k}(t_f) \nonumber\\
v_k(t_f) &=& u_k^0 b_{2k}(t_f) + v_k^0 b_{1k}(t_f) \label{uveq}
\end{eqnarray}
The total defect formation probability can now be obtained using
Eqs.\ \ref{uveq} and \ref{evol1}
\begin{eqnarray}
p_k &=& |\gamma_k|^2 = |b_{2k}(t_f)|^2= 4 p_k(1-p_k)
\sin^2(\phi_k^{\rm
st}) \nonumber\\
\phi_k^{\rm st} &=& \phi_k + \xi_k(t_2,t_1),\label{pfinal}
\end{eqnarray}
which is Eq.\ 18 of the main text. Here we have used Eq.\ 14 of the
main text for definition of $\gamma_k$. We note that the frequency
dependence of the Stuckelberg phase arises out of the dependence of
$\phi_k^{\rm st}$ on $\delta_k$ and $\xi_k$, both of which depend on
$\omega_0$ through $\lambda(t)$. The details of this dependence is
discussed in details in the main text.


\begin{thebibliography}{99}

\bibitem{rev1}A. Polkovnikov, K. Sengupta, A. Silva, and M. Vengalattore,
Rev. Mod. Phys., {\bf 83}, 863 (2011).

\bibitem{rev2} J. Dziarmaga, Adv. Phys. {\bf 59}, 1063 (2010).

\bibitem{rev3} A. Dutta, U. Divakaran, D. Sen, B. K. Chakrabarti, T. F.
Rosenbaum, and G. Aeppli, arXiv:1012.0653 (unpublished)

\bibitem{isingref} S. Sachdev, K. Sengupta, and S.M. Girvin, \prb {\bf 66}, 075128
(2002).

\bibitem{boseref} M. Greiner, O. Mandel, T. Esslinger, T. W. Hansch, and I.
Bloch, Nature {\bf 415}, 39 (2002).

\bibitem{bloch1}  I. Bloch, J. Dalibard, and W. Zwerge, Rev. Mod. Phys. {\bf 80}, 885
(2008).

\bibitem{bakr1} W. S. Bakr, A. Peng, M. E. Tai, R. Ma, J. Simon, J. I.
Gillen, S. Folling, L. Pollet, and M. Greiner, Science {\bf 329},
547 (2010).

\bibitem{venga1}  Y. S. Patil, L. M. Aycock, S. Chakram, M. Vengalattore, arXiv:1404.5583 (unpublished).

\bibitem{sengupta1} K. Sengupta, S. Powell and S. Sachdev, \pra {\bf 69}, 053616
(2004).

\bibitem{calabrese1} P. Calabrese and J. Cardy, J. of Stat. Mech.: Theor. and
Experiment {\bf 2005}, P04010 (2005); {\it ibid.}, \prl {\bf 96},
136801 (2006).

\bibitem{anatoli1} T. W. B. Kibble, J. Phys. A {\bf 9}, 1387 (1976); W. H. Zurek,
Nature (London) {\bf 317}, 505 (1985); A. Polkovnikov, Phys. Rev. B
{\bf 72}, 161201(R) (2005).

\bibitem{sen1}D. Sen, K. Sengupta and S. Mondal, \prl {\bf 101}, 016806
(2008); S. Mondal, K. Sengupta, and D. Sen, \prb {\bf 79}, 045128
(2009).

\bibitem{sengupta2} K. Sengupta, D. Sen and S. Mondal, \prl {\bf 100}, 077204
(2008); S. Mondal, D. Sen, and K. Sengupta, \prb {\bf 78}, 045101
(2008).

\bibitem{dutta1}  U. Divakaran, A. Dutta and D. Sen, Phys. Rev. B {\bf 78},
144301 (2008).

\bibitem{das1} A. Das, \prb {\bf 82}, 172402 (2010); S Bhattacharyya, A Das, and
S Dasgupta, \prb {\bf 86}, 054410 (2012).

\bibitem{sengupta3} S. Mondal, D. Pekker, and K. Sengupta, Europhys.
Lett. {\bf 100}, 60007 (2011).

\bibitem{das2} A. Lazarides, A. Das, and R. Moessner, Phys. Rev. Lett. {\bf 112}, 150401
(2014).

\bibitem{hugo1} H. Touchette, R. Leidl and A. K. Hartmann (eds),
{\it Modern Computational Science 11: Lecture Notes from the 3rd
International Oldenburg Summer School, BIS-Verlag der Carl von
Ossietzky Universitat Oldenburg}, (2011).

\bibitem{hugo2} H. Touchette, Phys. Rep. {\bf 478}, 1 (2009).

\bibitem{silva1} A. Silva, \prl {\bf 101}, 120603 (2008); A. Gambassi
and A. Silva, \prl {\bf 109}, 250602 (2012); S. Sotiriadis, A.
Gambassi, and A. Silva, \pre {\bf 87}, 052129 (2013).

\bibitem{silva2} P. Smacchia and A. Silva, \pre {\bf 88}, 042109
(2013).

\bibitem{heyl1} M. Heyl and S. Keherin, \prl {\bf 108}, 190601
(2012).

\bibitem{esposito1} G. Verley, C. Van den Broeck, and M.
Esposito, \pre {\bf 88} 032137 (2013).


\bibitem{comment0} Note that the equality in the last relation holds only if
$|\psi(t)\rangle$ has a perfect overlap with the initial ground
state, in which case $f(u)=0$ for all $u$ leading to $P(w) \sim
\delta(w)$; a situation very close to this happens if the system
exhibits perfect dynamic freezing \cite{das1,sengupta3}.

\bibitem{comment1} The most general form of the Gartner-Ellis
theorem  yields $I({\bar w})= -{\rm Inf}[u{\bar w}-f(u)]$, where
${\rm Inf}$ stands for Infimum \cite{hugo1}. However, if $f'[u]$
exists, this defination of $I({\bar w})$ coincides with the one
given in Eq.\ \ref{gerate}.

\bibitem{comment2} The solution is obtained by setting ${\bar w}=0$ and
then taking $u[{\bar w}] \to \infty$ so that ${\bar w}u[{\bar w}]
\to 0$.

\bibitem{intref} M. M. Rams and B. Damski, \prl {\bf 106}, 055701
(2011); H-Q Zhou, R. Orus, and G. Vidal, \prl {\bf 100}, 080601
(2008); H-Q Zhou and J. P. Barjaktarevic, J. Phys. A, {\bf 41},
412001 (2008).

\bibitem{comment00} See supplementary material for a more detailed
derivation.

\bibitem{markos1} M. Rigol, B. S. Shastry, and S. Haas
\prb {\bf 80}, 094529 (2009); C. de Grandi, V. Gritsev, and A.
Polkovnikov, \prb {\bf 81}, 012303 (2010).

\bibitem{commentfi} See supplementary materials for a detailed proof of this
statement.

\bibitem{commentex} As explained in the supplementary material,
$\sum_j p'_j \ge 0$ and is independent of ${\mathcal M}$ in the
${\mathcal M} \to \infty$ limit. Also, the excitations represented
by $p'_j$, in contrast to those for integrable models, need not be
local quasiparticles, but could be, for example, non-local objects
such as defects, vortices or other collective modes.

\bibitem{isingbook} See for example, S. Sachdev, {\it Quantum Phase Transitions}
(Cambridge University Press, Cambridge, England, 1999).

\bibitem{commentkit} See supplementary material for a discussion
of the equilibrium phases of these spin models.


\bibitem{kitaevref} A. Kitaev, Ann. Phys. (N.Y.) {\bf 321}, 2 (2006);
X.-Y. Feng, G.-M. Zhang, and T. Xiang, Phys. Rev. Lett. {\bf 98},
087204 (2007).


\bibitem{nori1} S.N. Shevchenko, S. Ashhab, F. Nori, Phys. Rept.
{\bf 492}, 1 (2010).

\bibitem{commentstu1} See supplementary materials for an
introduction to the Stuckelberg interference phenomenon and a sketch
of the derivation of $p_k$ within adiabatic-impulse approximation.

\bibitem{comment3} It turns out that for a periodic protocol which
constitutes a series of step functions, the adiabatic-impulse
approximation becomes exact.

\bibitem{prop1} R. Dorner, S. R. Clark, L. Heaney, R. Fazio, J. Goold, and V.
Vedral, \prl {\bf 110}, 230601 (2010).

\bibitem{prop2} L. Mazzola, G. De Chiara, and M. Paternostro, \prl {\bf 110}, 230602 (2010).

\bibitem{exp1}T. Batalhão, A. M. Souza, L. Mazzola, R. Auccaise, R. S. Sarthour,
I. S. Oliveira, J. Goold, G. De Chiara, M. Paternostro, and R. M.
Serra, arXiv:1308.3241 (unpublished).

\bibitem{hegde1} S. Hegde, H. Katiyar, T. S. Mahesh, and A. Das,
arXiv:1307.8219 (unpublished).

\bibitem{commentf} For example, for few-spin system, $n_d$ can be read
off from the number of spins deviating from the ground state
configuration; for a two-level system it involves measuring
occupation probability of the upper level.

\end{thebibliography}
\end{document}